\documentclass[onecolumn,usenatbib]{mn2e}
\usepackage[usenames,dvips]{color}
\usepackage{graphicx}
\usepackage{pstricks}
\usepackage{pst-node}
\usepackage{pst-tree}

\newcommand{\Fewbody}{{\em Fewbody\/}}
\newcommand{\GLStarView}{{\em GLStarView\/}}
\newcommand{\Starlab}{{\em Starlab\/}}

\title[Stellar collisions during binary--binary and binary--single star interactions]
      {Stellar collisions during binary--binary and binary--single star interactions}

\author[J.~M.~Fregeau, P.~Cheung, S.~F.~Portegies Zwart, \& F.~A.~Rasio]
       {J.~M.~Fregeau$^1$\thanks{E-mail: fregeau@mit.edu},
	 P.~Cheung$^2$,
         S.~F.~Portegies Zwart$^3$,
         and F.~A.~Rasio$^4$\\
	 $^1$Dept.\ of Physics, MIT, Cambridge, MA 02139\\
	 $^2$Dept.\ of Mathematics, Stanford University, Stanford, CA 94305\\
	 $^3$Astronomical Institute ``Anton Pannekoek'' and Section Computational Science, \\
	 ~~University of Amsterdam, Amsterdam, the Netherlands\\
	 $^4$Dept.\ of Physics and Astronomy, Northwestern University, Evanston, IL 60208}
       
\begin{document}

\date{Accepted for publication in MNRAS}

\pubyear{2004}

\maketitle

\begin{abstract}
Physical collisions between stars occur frequently in
dense star clusters, either via close encounters between two single stars, 
or during strong dynamical interactions involving binary stars.  Here we study stellar
collisions that occur during binary--single and binary--binary interactions, by performing
numerical scattering experiments.  Our results include cross
sections, branching ratios, and sample distributions of parameters for various outcomes.
For interactions of hard binaries containing main-sequence stars, we find that the 
normalized cross section for at least one collision
to occur (between any two of the four stars involved) is essentially unity, and that the 
probability of collisions involving more than two stars is significant.  Hydrodynamic 
calculations have shown that the effective 
radius of a collision product can be 2--30 times larger than the normal main-sequence
radius for a star of the same total mass.  We study the effect of this expansion,
and find that it increases the probability of further collisions considerably.
We discuss these results in the context of recent observations of blue stragglers in globular
clusters with masses exceeding twice the main-sequence turnoff mass.
We also present \Fewbody, a new, freely available numerical toolkit for simulating small-$N$
gravitational dynamics that is particularly suited to performing scattering experiments.
\end{abstract}


\section{Introduction}

Close encounters and direct physical collisions between stars occur frequently in
globular clusters.  For a star in a dense cluster core, the typical collision time 
can be comparable to the cluster lifetime, implying that essentially all
stars could have been affected by collisions \citep{1976ApL....17...87H}.
Even in moderately dense clusters, collisions can happen frequently during resonant 
interactions involving primordial binaries 
\citep{1983Natur.301..587H,1989AJ.....98..217L,1993ApJ...415..631S,1995MNRAS.276..876D,1995ApJS...99..609S,1996MNRAS.281..830B}.
In open clusters with significant
binary fractions ($\sim 10$\% or more), mergers may occur more often
through binary--binary interactions than through single--single collisions and
binary--single interactions combined \citep{1991AJ....102..994L}.
Collisions involving more than two stars can be quite common during
binary--single and binary--binary interactions, since the product of a first collision
between two stars expands adiabatically following shock heating,
and therefore has a larger cross section for subsequent collisions
with the remaining star(s).

Collisions and binary interactions strongly affect the dynamical evolution of
globular clusters.  The formation of more massive objects through mergers tends 
to accelerate core collapse, shortening cluster lifetimes.  On the other hand, mass loss from evolving 
collision products can indirectly heat the cluster core, thereby postponing core collapse.
The realization during the 1990s that primordial binaries are
present in globular clusters in dynamically significant numbers has completely
changed our theoretical perspective on these systems 
\citep{1989Natur.339...40G,1995ApJS...99..609S,ivanova2003}.
Most importantly, 
dynamical interactions of hard primordial binaries with single stars and other binaries 
are thought to be the primary mechanism for supporting globular clusters against core collapse
\citep{1990ApJ...362..522M,1991ApJ...372..111M,1991ApJ...370..567G,1992PASP..104..981H,1992MNRAS.257..513H,1994ApJ...427..793M,2001ibsp.conf..387R,2003ApJ...593..772F,2003MNRAS.343..781G}.
Observational evidence for the existence of primordial
binaries in globular clusters is now well established 
\citep{1992PASP..104..981H,1994ApJS...90...83C,1997ApJ...474..701R}.
Recent {\em Hubble Space Telescope\/} ({\em HST\/}) observations have provided direct 
constraints on the primordial binary 
fractions in many clusters.  For example, the observation of a broadened main 
sequence in NGC 6752, based on {\em HST\/}-WFPC2 images, suggests that the binary
fraction is probably in the range 15\%--40\% in the inner core
\citep{1997ApJ...474..701R}.  Using a similar method, \citet{2002AJ....123.1509B} 
find that the binary fraction in the inner region of NGC~288 is probably between
10\% and 20\%, and less than 10\% in the outer region.  Observations of eclipsing binaries and BY Draconis
stars in 47~Tuc yield an estimate of $\sim13$\% for the core binary fraction 
\citep{2001ApJ...559.1060A}, although a recent reinterpretation of the observations
in combination with new theoretical results
suggests that this number might be closer to $\sim5$\% \citep{ivanova2003}.
Using {\em HST\/}-WFPC2, \citet{1999AAS...195.7602B} derive an upper limit of only $\sim3$\% on the binary 
fraction in the core of NGC 6397.


In this paper, we focus on interactions involving main-sequence (hereafter MS) stars,
and the production of blue stragglers (hereafter BSs). BS stars 
appear along an extension of the MS blue-ward of the turnoff point in the colour--magnitude 
diagram (CMD) of a star cluster.  All observations suggest that they are massive MS stars 
formed through mergers of two (or more) lower-mass stars. For example, \citet{1998ApJ...507..818G}
have demonstrated that the masses estimated from the pulsation frequencies of four oscillating 
BSs in 47~Tuc are consistent with their positions in the CMD.  Indirect 
measurements of BS masses yield values of up to four times the MS turnoff mass, although
the uncertainties are significant \citep{1971PASP...83..638B,1971PASP...83..768S,1992IAUS..151..473M}.
More recent spectroscopic measurements yield much more precise masses,
with one BS in 47~Tuc about twice the MS turnoff mass \citep{1997ApJ...489L..59S}, and
two in NGC~6397 more than twice the MS turnoff mass \citep{2000AAS...196.4106S}.

Mergers of MS stars can occur in at least two different ways: via physical collisions, 
or through the coalescence of two stars in a close binary system 
\citep{1989AJ.....98..217L,1993blst.conf....3L,1993PASP..105.1081S,1995ApJ...439..705B}.
Direct evidence for binary progenitors has been found in the form of contact (W~UMa type) 
binaries among BSs in many globular clusters \citep{2000AJ....120..319R}, including
low-density globular clusters such as NGC 288 \citep{2002AJ....123.1509B}, NGC 5466 
\citep{1990AJ....100..469M}, and M71 \citep{1994AJ....108.1810Y}, as well as in many 
open clusters \citep[e.g.,][]{1995A&A...295..101J,1993MNRAS.265...34K,1994AJ....108.1828M}.
At the same time, strong indication for a collisional origin
comes from detections by {\em HST\/} of large numbers of bright BSs
concentrated in the cores of some of the densest clusters,
such as M15 \citep{1994ApJ...422..597D,1994AJ....107.1745Y,1996AJ....111..267G},
M30 \citep{1994ApJ...435L..59Y,1998AJ....116.1757G}, 
NGC 6397 \citep{1994A&A...287..769B},
NGC 6624 \citep{1995AJ....109..639S}, and M80 \citep{1999ApJ...522..983F}. 
High-resolution {\em HST\/} images reveal that the central density profiles in many of these clusters
steadily increase down to a radius of $\sim 0.1\,$pc, with no signs of 
flattening.  Direct stellar collisions should be extremely frequent 
in such high density environments.

Evidence for the greater importance of binary interactions over direct
collisions of single stars for producing BSs in some globular clusters can be 
found in a lack of correlation between 
BS specific frequency and cluster central collision 
rate \citep{2003nhgc.conf..296D,2003ApJ...588..464F}.  More direct evidence 
comes from the BS S1082 in the open cluster M67, which is
part of a wide hierarchical triple system \citep{2003AJ....125..810S}.
The most natural formation mechanism is via a binary--binary interaction.
There is further evidence in the radial distributions of BSs in clusters.  
{\em HST\/} observations, in combination with ground-based studies, have revealed
that the radial distributions of BSs in the clusters M3 and 47~Tuc
are bimodal -- peaked in the core, decreasing at intermediate radii, and rising
again at larger radii \citep{2004ApJ...603..127F,1993AJ....106.2324F,1997A&A...324..915F}.
The most plausible explanation is that the BSs at larger radii were
formed through binary interactions in the cluster core and ejected to larger radii \citep{1994ApJ...431L.115S}.

In this paper, we perform numerical scattering experiments to study stellar collisions that occur 
during binary interactions.  One approach for attacking the problem is to perform
a full globular cluster simulation, taking into account every relevant physical process, including
stellar dynamics, stellar evolution, and hydrodynamics.  This approach is enticing in 
its depth, but would certainly yield results with a complicated dependence on the input parameters
and physics that would be difficult to disentangle.  A simpler approach is to study 
in detail the scattering interactions that occur between binaries and single stars or
other binaries.  This approach isolates the relevant
physics and produces results that are easier to interpret.  Furthermore,
the cross sections tabulated will be useful for future analytical and numerical
calculations of cluster evolution and interaction rates.
For a discussion of the interplay between globular cluster dynamics and stellar
collisions, see, e.g., \citet{2001MNRAS.323..630H}.  For dense globular cluster cores, 
merger rates via binary stellar evolution can be significantly enhanced by dynamical 
interactions \citep{ivanova2003}.

Our paper is organized as follows.  In \S~2 we summarize previous theoretical
work on stellar collisions in binary interactions.  In \S~3 we describe our
numerical method, and introduce the two numerical codes used.  In \S~4 we test
the validity of our numerical method by comparing with previous results.
In \S~5 we present a systematic study of the dependence of the collision cross section in
binary--single and binary--binary interactions on several physically relevant 
parameters.  In \S~6 we consider binaries with parameters characteristic of
those found in globular clusters, and study the properties of the resulting binaries
and triples containing collision products.  Finally, in \S~7 we summarize and conclude.

\section{Previous Work}

There now exists a very large body of numerical work on binary--single and, to a lesser extent,
binary--binary interactions \citep[see, e.g.,][for an overview and references]{2003gmbp.book.....H}.
\citet{1983ApJ...268..319H} performed one of the most extensive early studies
of binary--single star scattering for the equal-mass, point-particle case.
\citet{1983MNRAS.203.1107M} performed the first systematic studies
of binary--binary interactions in the point-particle limit, first for the case of
equal energy binaries, and later for unequal energies \citep{1984MNRAS.207..115M}.
He also studied the energy generated in binary--binary interactions in the context
of the evolution of globular clusters \citep{1983MNRAS.205..733M,1984MNRAS.208...75M}.
Most numerical scattering experiments have been performed in the point-mass limit, neglecting altogether
the effects of the finite size of stars.  However, as we summarize below,
there are a number of studies that apply approximate
prescriptions for dissipative effects and collisions {\em post facto\/} to 
numerical integrations performed in the point-mass limit but
in which pairwise closest approach distances were recorded.
There is also one study in which collisions are treated {\em in situ\/} in a 
simplified manner, and several that perform full smoothed particle hydrodynamics 
(SPH) simulations of multiple-star interactions (see below).

\citet{1983AJ.....88.1420H} was the first to study distances of closest
approach between stars in both binary--single and binary--binary interactions. He found that roughly 
40\% of binary--binary encounters in a typical globular cluster core will lead to a physical 
collision between two stars.  \citet{1984ApJ...282..466K} considered the evolution of a compact
binary in a globular cluster core subject to perturbations by single field stars, and found that 
an induced merger or collision between two stars in a binary--single interaction is likely.
\citet{1985ApJ...298..502H} applied a single-parameter, ``fully inelastic sphere''
collision model after the fact to a large number of
binary--single interactions for which distances of closest approach were recorded, and
calculated merger rates.  
\citet{1986ApJ...306..552M}
applied simple prescriptions for the dissipative effects of gravitational radiation,
tidal interactions, and physical contact between stars after the fact to a large number
of binary--single interactions involving tight binaries.  He found that dissipative
effects reduce binary heating efficiency in cluster cores by roughly an order of magnitude
over that obtained in the point-mass limit.  He also found that the most likely outcome
of a binary--single interaction involving a tight binary is the coalescence of at least two
of the stars.  
This work was carried further by \citet{1997A&A...328..143P,1997A&A...328..130P},
who included the effects of binary stellar evolution on binary--single interactions.
They found that about 20 per cent of encounters between a primordial binary and a cluster
star result in collisions, while almost 60 per cent of encounters with tidal-capture
binaries lead to collisions.
\citet{1989AJ.....98..217L} performed a small number of binary--binary 
interactions and recorded close approach distances and calculated ejection speeds of 
collision products.
\citet{1991AJ....102..994L} performed the first set of binary--single and binary--binary
interactions in which stars were allowed to merge {\em during\/} the numerical
integration.  They studied the rate of production of BSs in clusters,
and performed the first, simplified ``population synthesis'' study of BSs
in clusters.  \citet{1991AJ....102..704H,1992AJ....103.1955H} considered stars with a range of masses
exchanging into binaries, and found the distance of closest approach to be roughly a 
constant fraction of binary semimajor axis independent of intruder mass, over a wide 
range of mass ratios.  \citet{1990ApJ...349..150C} performed full SPH simulations
of binary--single interactions and showed directly the importance of taking into account
the non-zero size of stars.  
\citet{1991ApJ...378..637G} and \citet{1993ApJ...411..285D,1994ApJ...424..870D} performed 
sets of binary--single
and binary--binary interactions with tight binaries in the point-mass limit, and selected a 
handful to run with a full SPH code.  They found that multiple
mergers are common.  \citet{1996MNRAS.281..830B} performed a large number of binary--binary
interactions and presented a survey of close approach cross sections for several
sets of physically relevant binary parameters.  They also calculated outcome frequencies,
studied the properties of the interaction products, and used their results
in analytical calculations of interaction rates in globular cluster cores.  
More recently, \citet{2003MNRAS.343..781G} have incorporated into their Monte-Carlo
globular cluster evolution code Aarseth's {\em NBODY} for performing direct integrations of 
binary interactions.  By storing the results of binary interactions that occur during
cluster evolution, they have calculated close approach cross sections, and a few 
differential cross sections.

\section{Numerical Method}

The scattering experiments presented in this paper were performed primarily using \Fewbody,
a new numerical toolkit designed for simulating small-$N$ gravitational dynamics,
which we describe below.  
In some cases we also use the scattering facilities of the \Starlab\ software
environment \citep{2001MNRAS.321..199P}.  \Starlab\ was used mainly to compare with \Fewbody, 
but in cases where \Starlab\ data were compiled before \Fewbody\ was written, \Starlab\
results were used.  In particular, all calculations in \S~\ref{sec:realistic} were performed
with \Starlab.

\subsection{Setup}
We label the two objects in a scattering experiment 0 and 1.  In the case 
of binary--single scattering, 0 is the binary and 1 is the single star.  In the case of binary--binary
scattering, 0 and 1 are each binaries.  We use the same system of labelling for each binary,
so the members of binary $i$ are labelled $i0$ and $i1$.  There are several parameters 
required to uniquely specify a binary--single or binary--binary scattering experiment.
To describe the initial hyperbolic (or parabolic) orbit between objects 0 and 1, 
one needs to specify the relative velocity at infinity $v_\infty$, impact parameter $b$, and masses
$m_0$ and $m_1$.  To describe the internal properties of each object, one needs to specify
the semimajor axes $a_i$, eccentricities $e_i$, individual masses $m_{ij}$, and stellar
radii $R_{ij}$.  There are also several phase and orientation angles required for each binary:
the orientation of the binary angular momentum vector relative to the angular momentum vector
describing the orbit between 0 and 1, given by the polar angle $\theta$ and the azimuthal angle $\phi$; 
the angle $\omega$ between the binary Runge-Lenz vector and some fiducial vector perpendicular to the 
binary angular momentum vector (e.g., the 
cross product of the binary angular momentum and the 0-1 angular momentum); and $\eta$,
the mean anomaly of the binary.  For all the scattering experiments presented in this paper, these
phase and orientation angles are chosen randomly, so that the cross sections calculated represent
averages over these quantities.  In detail, these angles are given by
$\theta=\cos^{-1}(2X-1)$, $\phi = 2\pi X$, $\omega = 2\pi X$, and $\eta = 2\pi X$,
where $X$ is a uniform deviate in the range $[0,1)$.
In addition, unless otherwise noted, the eccentricity of each
binary is chosen from a thermal distribution \citep{1919MNRAS..79..408J}
truncated at large $e$ such that there is no contact binary.  In each scattering experiment,
numerical integration is started at the point at which the tidal perturbation 
($F_{\rm tid}/F_{\rm rel}$) on a binary in the system reaches $\delta$ (see \S~\ref{sec:isolation}).

It is customary to specify the relative velocity at infinity
in terms of the critical velocity, $v_c$, defined such that the total energy of the
binary--single or binary--binary system is zero.  For $v_\infty > v_c$ the total energy of the
system is positive, and full ionization is possible.  That is, a possible outcome of the scattering
experiment is that each star leaves the system unbound from any other with positive velocity
at infinity.  For $v_\infty < v_c$, the total energy
of system is negative, and the encounters are likely to be resonant, with all stars involved
remaining in a small volume for many dynamical time-scales.  Defining the total mass $M=m_0+m_1$ 
and reduced mass $\mu=m_0m_1/M$, the critical velocity is
\begin{equation}\label{eq:vcbs}
  v_c = \left[\frac{G}{\mu}\left(\frac{m_{00}m_{01}}{a_0}\right)\right]^{1/2}
\end{equation}
for the binary--single case, and
\begin{equation}\label{eq:vcbb}
  v_c = \left[\frac{G}{\mu}\left(\frac{m_{00}m_{01}}{a_0} + \frac{m_{10}m_{11}}{a_1}\right)\right]^{1/2}
\end{equation}
for the binary--binary case.  The cross section for outcome X is obtained by performing many scattering
experiments out to a maximum impact parameter $b_{\rm max}$ and calculating
\begin{equation}
  \sigma_X = \pi b_{\rm max}^2 \frac{N_X}{N} \, ,
\end{equation}
where $N_X$ is the number of experiments that have outcome X, and $N$ is the total number of
scattering experiments performed.  In all cases the maximum impact parameter was chosen
large enough to ensure that the full region of interest was sampled.  In other words, for 
$b>b_{\rm max}$, all interactions are fly-by's in which each binary is only weakly tidally
perturbed during the interaction.  For calculations performed with \Fewbody,
$b_{\rm max}$ was chosen to correspond to a pericentre distance of $r_p=5(a_0+a_1)$ in the
binary--binary case, and $r_p=5a$ in the binary--single case.  For this value of pericentre
distance, the binary eccentricity induced in the fly-by is quite small
\citep[$\delta e \ll 1$ for initially circular binaries, and $\delta e/e \ll 1$ for non-circular binaries; see][]{1995ApJ...445L.133R,1996MNRAS.282.1064H}.
For calculations performed with \Starlab, $b_{\rm max}$ was chosen automatically by using successively
larger impact parameter annuli until no relevant outcomes were found \citep{1996ApJ...467..348M}.
The uncertainty in the cross section is calculated assuming Poisson counting statistics, so that
\begin{equation}
  \Delta\sigma_X = \pi b_{\rm max}^2 \frac{\sqrt{N_X}}{N} \, .
\end{equation}
In principle, it is necessary to include scattering experiments that result in unresolved
outcomes in this uncertainty \citep[see, e.g.,][]{1983ApJ...268..319H}.  
However, in practice we find that the number of unresolved outcomes is small, and does
not significantly contribute to $\Delta\sigma_X$.

\subsection{Possible Outcomes}
The possible outcomes of binary--single and binary--binary scattering interactions are listed
in Tables \ref{tab:bsoutcomes} and \ref{tab:bboutcomes}, ordered by the number of collisions,
$n_{\rm coll}$.  Stars are represented as filled
circles, brackets enclose two objects that are bound to each other in a binary, and colons represent
physical collision products.  In Table \ref{tab:bboutcomes} we also list the abbreviations
used in the paper to refer to certain outcomes.
When there are no collisions (as is the case in the point-mass
limit), the number of possible outcomes is small, as shown in the $n_{\rm coll}=0$ rows in each table.
However, when one considers stars with non-zero radius and allows for the possibility of collisions
and subsequent mergers, the total number of outcomes becomes large.  Assuming 
indistinguishable stars, there are
six possible outcomes for the binary--single case, and 16 for the binary--binary case.  These numbers
are evidently increased for distinguishable stars.  The software used in this paper distinguishes
among all possible outcomes.

\subsection{\Fewbody}\label{sec:fewbody}
\Fewbody\ is a new numerical toolkit for simulating small-$N$ gravitational dynamics.  
It is a general $N$-body dynamics code, although it was written 
for the purpose of performing scattering experiments, and therefore has several
features that make it well-suited for this purpose.  It can
be described succinctly in terms of its key elements.

\subsubsection{Adaptive Integration and Regularization}
At its core, \Fewbody\ uses the 8th-order Runge-Kutta Prince-Dormand integration
method with 9th-order error estimate and adaptive timestep to advance the 
$N$-body system forward in time.  It integrates the usual formulation of the
$N$-body equations in configuration space, but allows for the option of global pairwise 
Kustaanheimo-Stiefel (K-S) regularization \citep{1974CeMec..10..217H,1985MNRAS.215..171M}.
Global regularization is a coordinate transformation that removes all 
singularities from the $N$-body equations, making
the integration of close approaches, and even collision orbits, much more accurate.
It is well-suited for small-$N$ dynamics, since it requires the integration
of $\sim N^2$ separations instead of $N$ positions, and becomes prohibitively
computationally expensive for $N \ga 10$.  Although it should in principle make 
numerical integration more accurate, it was found that the adaptive timestep 
algorithm alone performed as well as
global regularization, in terms of the computational time required 
for a specified level of energy and angular momentum accuracy.  The use
of regularization requires extra effort to detect physical collisions, 
since, with regularization, pericentre is not necessarily resolved by
the integrator.  For the sake of simplicity, we have chosen not to implement 
the appropriate technique for detecting collisions with regularization 
\citep[see \S~9.8 of][]{aarsethbook}.  Furthermore, physical collisions naturally 
soften the singularities in the non-regularized $N$-body equations by making them 
physically inaccessible.  Regularization was therefore only used to test 
calculations made in the point-mass limit.  For all other calculations the 
non-regularized integration routine was used.

\subsubsection{Classification}\label{sec:classification}
\begin{figure}
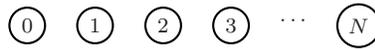

  \begin{center}
    \pstree[levelsep=40pt]{\Tcircle{0}}{}
    \quad
    \pstree[levelsep=40pt]{\Tcircle{1}}{}
    \quad
    \pstree[levelsep=40pt]{\Tcircle{2}}{}
    \quad
    \pstree[levelsep=40pt]{\Tcircle{3}}{}
    \quad
    $\cdots$
    \quad
    \pstree[levelsep=40pt]{\Tcircle{$N$}}{}
    \caption{\label{fig:flattree}Schematic representation of the binary-tree algorithm used
      by \Fewbody.  A circle containing a number represents a star.  The set of binary trees 
      is shown flattened here, as it is before processing, 
      so that each star is the top-level node of a one-node tree.}
  \end{center}
\end{figure}

\begin{figure}
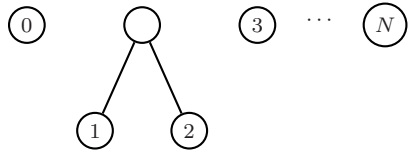

  \begin{center}
    \pstree[levelsep=40pt]{\Tcircle{0}}{}
    \quad
    \pstree[levelsep=40pt]{\Tcircle{\phantom{0}}}{
      \Tcircle{1}
      \Tcircle{2}
    }
    \quad
    \pstree[levelsep=40pt]{\Tcircle{3}}{}
    \quad
    $\cdots$
    \quad
    \pstree[levelsep=40pt]{\Tcircle{$N$}}{}
    \caption{\label{fig:firstpasstree}Schematic representation of the binary-tree algorithm used
      by \Fewbody.  A circle containing a number represents a star, while an empty circle represents
      a general parent node.  The set of binary trees is shown after the first stage of processing,
      with stars 1 and 2 replaced by their parent node, which contains the dynamical information
      pertaining to the centre of mass of the 1-2 binary, as well as all phase and orientation
      information.  For classification, the replacement of stars 1 and 2 by their parent node simply
      means that they are bound to each other with the smallest semimajor axis.  For hierarchy
      isolation, it would also mean that the 1-2 binary is sufficiently weakly perturbed
      that it can be treated analytically, and is stable in the sense that its two members will
      not collide at pericentre.}
  \end{center}
\end{figure}

\Fewbody\ uses a binary tree algorithm to handle several aspects related to performing
scattering experiments.  Most importantly, it uses a binary tree algorithm to classify
the $N$-body system into a set of independently bound hierarchies.  For example, if the outcome
of a scattering experiment between two hierarchical triples is a hierarchical triple 
composed of binaries, \Fewbody\ will classify it accordingly.  \Fewbody\ creates the
set of binary trees iteratively, according to the following simple rules.  First, 
as shown in Figure \ref{fig:flattree}, any existing set of trees is flattened so that 
each star in the $N$-body system represents the top-level node of a one-node tree.  Next, 
as shown in Figure \ref{fig:firstpasstree}, the two top-level nodes that are bound to each
other with the smallest semimajor axis are replaced by a parent node containing all dynamical 
information about the centre of mass, as well as all information about the binary's orbit, 
including phase.  The previous step is repeated, as shown in Figure \ref{fig:secondpasstree}, 
until no top-level nodes are found to be bound to each other.  This algorithm is clearly
general in $N$.  The resulting set of binary trees is a unique classification of the
configuration of the $N$-body system.  As described below, the classification is used for determining when an
interaction is complete.  The binary tree algorithm is also used (with a slightly different set of 
rules for creating the trees) to make the numerical integration more efficient, as also described below.

\subsubsection{Stability}\label{sec:stability}
\Fewbody\ assesses the dynamical stability of gravitationally bound hierarchies in an approximate
way using the classification just described, and a simple analytical test.  There currently exists
only one reasonably accurate criterion for the dynamical stability of an $N>2$ gravitational system,
the approximate analytical criterion of \citet{2001MNRAS.321..398M} 
for the dynamical stability of hierarchical triples.  \Fewbody\ assesses the stability
of each binary tree by applying this criterion at each level in the tree.
For example, for a hierarchical quadruple system (which consists of a star in orbit
around a hierarchical triple -- shown as ``\null[[[$\bullet$ $\bullet$] $\bullet$] $\bullet$]''
in Table \ref{tab:bboutcomes}), 
it first applies the triple stability criterion to the inner triple, then applies 
it to the ``outer'' triple, treating the innermost binary as a single object.  For the
case of a hierarchical quadruple composed of two binaries 
(``\null[[$\bullet$ $\bullet$] [$\bullet$ $\bullet$]]'' in Table \ref{tab:bboutcomes}), 
\Fewbody\ uses the additional
correction factor presented in \S~4.2 of \citet{2001MNRAS.321..398M}.  The stability
of a hierarchical system as determined by this method is only approximate, but, in our
experience, seems to work reasonably well.  For the particular case of binary--binary scattering, 
hierarchical triples which appear to be stable are classified as unstable less than roughly
one percent of the time.

It should be noted that the stability assessed here is dynamical rather than secular, so, e.g., any
resonances that would destroy an otherwise stable hierarchical system are ignored.  Such
resonances are likely to be important in the more general context of the dynamics
of globular clusters and their constituent populations \citep[e.g.,][]{2002ApJ...576..894M,2000ApJ...535..385F}, 
but are beyond the scope of the present paper.  It should also be noted
that our use of binary trees prevents us from recognizing stable three-body systems 
which are not hierarchical, such as the stable ``figure eight'' orbit for three stars of 
comparable mass \citep{2000AnnMath.152.881C,2001NAMS...48..471M}.  However, the fraction
of strong binary--binary scattering encounters resulting in this configuration is likely to be
exceedingly small \citep{2000MNRAS.318L..61H}.

\subsubsection{Hierarchy Isolation}\label{sec:isolation}
\Fewbody\ also uses a binary tree algorithm to speed up numerical integration by 
isolating from the integrator certain tight binaries and hierarchies that are only 
very weakly perturbed, yet dominate the calculation by driving down the integration 
time-scale.  It does this by integrating only the top-level nodes (centres of mass) of
a set of binary trees created using the algorithm described in \S~\ref{sec:classification},
but subject to a slightly different rule-set.  Two top-level nodes can only be replaced by
their parent node if: (1) the binary
tree represented by the parent node is stable, (2) the tidal perturbation on the outer
binary of the tree (the two nodes below the top-level) at apocentre due to all other top-level
nodes in the system is less than a specified fraction, $\delta$, of the minimum force between
them ($F_{\rm tid}/F_{\rm rel} < \delta$ at apocentre), 
and (3) the evolution of the binary tree can be treated analytically.  
The relative force at apocentre is calculated simply as
\begin{equation}
  F_{\rm rel}=\frac{Gm_0m_1}{[a(1+e)]^2} \, ,
\end{equation}
where $m_0$ and $m_1$ are the masses of the members of the outer binary,
$a$ is the semimajor axis, and $e$ is the eccentricity.
The tidal force at apocentre is calculated simply as
\begin{equation}
  F_{\rm tid}=\sum_i \frac{2G(m_0+m_1)m_i}{r_i^3} a (1+e) \, ,
\end{equation}
where the sum is taken over all other top-level nodes in the system, $m_i$ is the mass
of the other top-level node, and $r_i$ is the distance to the other top-level node.  Note that this sum represents
the upper limit of the tidal force since it does not take into account relative
inclination between the binary and the other top-level nodes.

A binary (or hierarchy) that is isolated from the integrator in this way is treated numerically
again when its relative tidal perturbation exceeds $\delta$.  This is done by resuming the
integration from the previous step (when the hierarchy's tidal perturbation was less than
$\delta$) with the parent node replaced by its child nodes, and orbital phase advanced to the
current time.  In practice,
this algorithm isolates from the integrator mainly weakly-perturbed binaries, and 
a few extremely hierarchical triples in which the tidal perturbation on the inner binary due to
the outer member is very small.  For binary--single scattering, hierarchy isolation 
can speed up the integrations
by up to an order of magnitude on average.  For binary--binary scattering, especially when the 
two binaries have very disparate semimajor axes, and hence orbital time-scales, this algorithm
can speed up the integrations by a few orders of magnitude on average.  The quantity
$\delta$ plays the role of an integration tolerance parameter.  Larger values of $\delta$
allow hierarchies to be treated analytically more frequently, yielding faster calculations
but sacrificing energy accuracy.  Smaller values of $\delta$ yield better energy conservation
at the expense of computational speed.

\subsubsection{Calculation Termination}
\Fewbody\ uses the classification and stability assessment techniques outlined above,
in combination with a few simple rules to automatically terminate the integration of
scattering encounters when they are complete -- in other words, when the separately bound 
hierarchies comprising the system will no longer interact with each other or evolve 
internally.  Integration is terminated when: (1) each pair of top-level nodes has 
positive relative velocity, (2) the tidal perturbation ($F_{\rm tid}/F_{\rm rel}$) 
on the outer binary (the two nodes below the top-level) of each tree 
due to the other top-level nodes is smaller than $\delta$, (3) each tree is dynamically
stable (as defined in \S~\ref{sec:stability}), and (4) the $N$-body system composed of 
the top-level nodes has positive energy.  The last condition is required because it is 
possible for the members of an $N$-body system to be separately unbound and receding 
from each other, yet for the system as a whole to be bound.  Here $\delta$ again
plays the role of an accuracy parameter, with smaller $\delta$ yielding
more accurate outcome classifications.  Since the $N$-body problem is chaotic,
with initially neighbouring trajectories in phase space diverging exponentially,
the value of $\delta$ should play only a minor role in the statistical accuracy
of classifications of outcomes.

\subsubsection{Physical Collisions}
\Fewbody\ performs collisions between stars in the ``sticky star'' approximation.
In this approximation, stars are treated as rigid spheres with radii equal to their
stellar radii.  When two stars touch, they are merged with no mass lost, and with linear momentum
conserved.    (Tidal effects, which may significantly increase the collision rate for close 
encounters \citep[see, e.g.,][]{1986ApJ...306..552M}, are beyond the scope of this method, 
but may be approximated by larger initial effective stellar radii.)
The radius of the merger product is set to
\begin{equation}
  R_{\rm merger} = f_{\rm exp} (R_1+R_2) \, ,
\end{equation}
where $R_1$ and $R_2$ are the radii of the merging stars, and $f_{\rm exp}$ is an
expansion factor.  To determine a reasonable value for $f_{\rm exp}$, one must consider
the relevant time-scales involved.  The characteristic time-scale of a typical binary 
scattering encounter in a globular cluster core is between $\sim 10\,\mbox{yr}$ for a fly-by
and $\la 10^4\,\mbox{yr}$ for a resonant encounter, while the 
thermal time-scale of a $\sim 1\,M_{\sun}$ MS star is $\sim 10^7\,\mbox{yr}$.  Therefore, it is 
invalid to treat merger products as rejuvenated (``reborn'') MS stars ($f_{\rm exp}=1$) 
during scattering encounters.  The hydrodynamical time-scale is
$\sim 1\,\mbox{hr}$, so it is more accurate to treat merger products as hydrodynamically
settled.  SPH simulations show that $f_{\rm exp}$ 
should be in the range 2--30, depending on the relative orientations of the two stars
before collision \citep{2003MNRAS.345..762L}.  These simulations also show that the amount
of mass lost in the types of collisions characteristic of globular clusters
is typically of order 1\%, so our assumption of zero mass loss is a reasonable first approximation.

Collision products are likely to have significant rotation and be non-spherical.
Furthermore, it is not clear that the value of the expansion parameter for the merger of
two pristine MS stars should be the same as that for mergers involving 
collision products.  Thus $f_{\rm exp}$ should be considered an effective quantity,
averaged over many collisions.  A more realistic approach that adopts several 
separate parameters is in principle possible, but beyond the scope of the current
paper.

\subsubsection{General Availability}
\Fewbody\ is freely available for download on the 
web\footnote{See http://www.mit.edu/\~{}fregeau, or search the web for ``Fewbody''.},
licensed under the GNU General Public License (GPL).  It contains a collection
of command line utilities that can be used to perform individual scattering and 
$N$-body interactions, but is more generally a library of functions that
can be used from within other codes.  Its facilities make it aptly suited for
performing scattering interactions from within larger numerical codes that, e.g.,
calculate cross sections, or evolve globular clusters via Monte-Carlo techniques.

Available along with \Fewbody, there is an OpenGL-based visualization tool
called \GLStarView\ that can be used to view $N$-body interactions as they are
being calculated by \Fewbody, in an immersive, 3-D environment.  \GLStarView\
has proven to be a valuable aid in developing our understanding and
physical intuition of binary interactions.

\subsection{\Starlab}\label{sec:starlab}
\Starlab\ is a collection of modular software tools designed to simulate the 
evolution of dense stellar systems and analyse the resulting data 
\citep[see][for a detailed description]{2001MNRAS.321..199P}. It is freely available 
on the web\footnote{See http://www.manybody.org.}.  It consists of a 
library of programs for performing stellar dynamics, stellar 
evolution, and hydrodynamics, together with a set of programs acting as bridges between them.  
They may be combined to study all aspects of the evolution of $N$-body systems.  
For this paper, we use the three-body scattering facility {\em scatter3\/} and 
the general $N$-body scattering facility {\em scatter\/} from version 3.5 of
\Starlab, along with {\em sigma3\/} and {\em sigma\/} for the automated calculation
of cross sections.

\section{Tests and Comparisons}

To assess the validity of calculations performed with \Fewbody, we have compared
the results of several scattering experiments with the results of previous studies.  For general
binary--single interactions, we have compared our results with those of \citet{1983ApJ...268..319H};  
for general binary--binary, \citet{1983MNRAS.203.1107M}; and
for detecting close approach distances, we have compared with binary--binary
calculations performed by \citet{1996MNRAS.281..830B}.  The scattering facilities
in \Starlab\ have been used extensively and tested thoroughly 
\citep[see, in particular,][]{gualandris2004}.  However, there has
only been one reported comparison between the three-body scattering routine and the
$N$-body routine in the literature \citep{gualandris2004}.  Below, we perform a new test and show 
that the two routines agree at a basic level.

\subsection{General binary--single comparison}

\citet{1983ApJ...268..319H} performed one of the most extensive early studies
of binary--single star scattering for the equal-mass, point-particle case.  
Figure~\ref{fig:hutcomp} shows a comparison of the results of $8\times 10^5$ 
scattering interactions calculated using \Fewbody\ with their Figure~5.  Plotted are the total
dimensionless cross sections ($\sigma/(\pi a^2)$, where $a$ is the binary semimajor axis)
for ionization (shown by star symbols) and exchange (triangles) as a function of 
$v_\infty/v_c$, for the equal-mass, zero eccentricity, point-particle case.  The dotted
lines represent the data from their figure (without error bars), while the straight
solid and dashed lines are the theoretically predicted cross sections for ionization and 
exchange from the same paper.  The agreement between the two is excellent, although it
appears that \citeauthor{1983ApJ...268..319H} systematically find a slightly larger 
cross section for ionization.  We note, however, that the two agree at roughly the
one-sigma level.

\subsection{General binary--binary comparison}

The first systematic study of binary--binary scattering was presented by 
\citet{1983MNRAS.203.1107M}.  He considered binaries with equal semimajor axes, and
stars of equal mass, in the point-particle limit.  We have chosen to compare with Table~5
of \citet{1983MNRAS.203.1107M}, which presents sets of scattering experiments performed
for several different values of $v_\infty/v_c$, with impact parameter chosen uniformly
in area out to the maximum impact parameter found to result in a strong interaction
(listed in his Table~3).  Only strong interactions were counted, and the 
eccentricities of the binaries were chosen from a thermal distribution.  It should be noted,
for the sake of completeness, that Mikkola characterised his encounters by their dimensionless energy
at infinity, $T_\infty$.  The relation between $v_\infty$ and $T_\infty$ is
$v_\infty/v_c = \sqrt{T_\infty}$.  Mikkola's
classification scheme is similar to \Fewbody's, the two primary differences
being: (1) the value of the tidal tolerance, $\delta$, used by Mikkola is $3\times 10^{-4}$, while the \Fewbody\
runs use $\delta=10^{-5}$; and (2) the criterion used to assess the dynamical stability of
triples is that of \citet{1974CeMec...9..465H}, a much less accurate stability
criterion than the \citet{2001MNRAS.321..398M} criterion used by \Fewbody.  It is 
therefore expected that the classification of \Fewbody\ is more accurate.
The binary--binary scattering encounters are classified into five different
outcomes.  The label ``undecided'' represents an encounter that was deemed to be unfinished
after a preset amount of computation time -- in other words, it could not be classified
into one of the four categories of ``exchange'', ``triple'', ``single ionization'', 
or ``full ionization''.  These four outcomes are described in the $n_{\rm coll}=0$ rows
of Table~\ref{tab:bboutcomes}.
Table~\ref{tab:mikkolacomp} compares results from \Fewbody\ with Mikkola's Table~5.  
The comparison is also shown graphically in Figure~\ref{fig:mikkolacomp}.

Several comments are in order.  Looking at the ``undecided'' column in Table
\ref{tab:mikkolacomp}, it is clear that \Fewbody\ resolves more encounters than Mikkola, 
yielding roughly half as many undecided encounters.  This is a result of both
the increased power of modern computers -- resonant encounters can be integrated longer,
and one can use smaller $\delta$ -- and the more accurate triple stability criterion 
available today.  In the next
column, labelled ``exchange'', it is clear that Mikkola finds many more exchange encounters
than \Fewbody.  This is thought to be primarily because in this column Mikkola's data
include strong interactions, which result not only in exchange, but also in preservation.  We have not
included this type of outcome in the \Fewbody\ results because it would have been cumbersome to
implement Mikkola's test for a strong interaction.  The next column, labelled
``triples'', shows that Mikkola regularly classifies more triples as stable than 
\Fewbody.  This results in fewer outcomes labelled as ``single ionization'',
since the test for single ionization occurs after that of triple stability in Mikkola's
code.  Full ionizations can only occur when the total energy of the system is greater than
or equal to zero ($v_\infty/v_c \geq 1$).  There is a large discrepancy
in the number of full ionizations
for $v_\infty/v_c=1.225$.  We are not quite sure of the underlying reason
for the discrepancy, but think it may be due to the tidal tolerance used, which differs
by more than an order of magnitude between the two methods.  Aside from the systematic 
discrepancies pointed out above, the two methods agree at a reasonable level, given the
differences between them.  This is especially clear from Figure~\ref{fig:mikkolacomp}.
For all outcomes except full ionization, the methods agree at roughly the two-sigma level 
(the uncertainties shown are one-sigma).

\subsection{Comparison for close-approach distances}

\citet{1996MNRAS.281..830B} presented a more recent and detailed
study of binary--binary interactions in the point-particle limit, in 
which close-approach distances were recorded and used to calculate cross
sections.  In the scattering experiment we have chosen for comparison,
each binary had equal semimajor axis ($a_0=a_1=a$) and zero eccentricity, and all stars
had equal mass.  The impact parameter was chosen uniformly in area out
to the maximum impact parameter given by $b_{\rm max}/a=C/v_\infty+D$, where
$C=5$, and $D=0.6$.
This expression for the impact parameter is an extension of that used by
\citet{1983ApJ...268..319H}, designed to sample strong
interactions adequately.  For each encounter, the minimum pairwise close approach
distance, $r_{\rm min}$, was recorded; and from the set, the cumulative cross section
calculated.  

Figure~\ref{fig:baconcomp} shows a comparison with their Figure~4.  The circles
with error bars represent \Fewbody\ data, while the solid-line broken power-law
is the best fit to the results obtained by \citet{1996MNRAS.281..830B}.
There is clearly a multiple-sigma discrepancy for $r_{\rm min}/a \la 0.01$.  
The discrepancy results from the lack of use by \citet{1996MNRAS.281..830B} of the appropriate algorithm
for detecting close approach distances with regularization \citep[\S~18.4 of ][]{aarsethbook}.
Sigurdsson has resurrected the original code, and performed a recalculation 
with smaller timesteps\footnote{See http://www.astro.psu.edu/users/steinn/4bod/index.html.}.  
The new result is shown by the dot-dash line.  The resulting cross section is closer 
to the \Fewbody\ result, yet still systematically smaller.

For comparison, we have performed the same calculation using \Starlab, 
shown by the dashed line.  The agreement between \Fewbody\ and \Starlab\ is
excellent.  The only discrepancy between the two occurs at $r_{\rm min}/a \sim 1$,
which represents the weak perturbation of binaries due to distant fly-by's.
This discrepancy is most likely due to the differing values of the tidal tolerance
used.  For the \Fewbody\ runs, the tidal tolerance was $\delta=10^{-5}$, while for
the \Starlab\ runs it was $\delta=10^{-6}$, causing \Starlab\ to numerically integrate
some weakly-perturbed binaries that \Fewbody\ treated analytically.  The result is
a slightly larger \Starlab\ cross section for $r_{\rm min}/a \sim 1$, as can be 
seen in the figure.  

We should note that the original calculation of
\citet{1996MNRAS.281..830B} was averaged over the range 
$0.125 \leq v_\infty/v_c \leq 0.25$, while all other results shown in Figure \ref{fig:baconcomp}
were calculated with $v_\infty/v_c=0.25$.  This cannot account for the 
discrepancy with the original calculation, since the inclusion of smaller
velocities at infinity will result in more resonant interactions, and hence smaller
distances of close approach.  We have performed calculations with 
$v_\infty/v_c=0.125$ and found that the cross section differs from that with
$v_\infty/v_c=0.25$ by no more than a few per cent.

Finally, we remark that the
error in the original calculation of \citet{1996MNRAS.281..830B} is only present
for small $r_{\rm min}$; many of the conclusions in their paper are not affected by
this error.

\subsection{Comparison between \Starlab's three-body and $N$-body scattering routines}

The scattering facilities in \Starlab\ have been used extensively and tested thoroughly
\citep{1996ApJ...467..348M,gualandris2004}.  However, there is only one reported
comparison between {\em scatter3\/}, the three-body scattering routine, and {\em scatter\/}, 
the $N$-body scattering routine, in the literature \citep{gualandris2004}.  
A simple test, tuned to suit the purposes
of this paper, is to compare the binaries containing merger products that result
from binary--single interactions with those from binary--binary interactions designed
to mimic binary--single interactions.  An obvious choice for the limiting-case 
binary--binary interaction is that in which one binary has an extremely small mass ratio.  
We performed binary--single runs
in which each star had mass $M_{\sun}$, radius $R_{\sun}$, the binary had
semimajor axis $1\,\mbox{AU}$ and $e=0$, and $v_\infty=10\,\mbox{km/s}$.  In the 
binary--binary runs, the binary mimicking the single star had a secondary 
of mass $10^{-5}\,M_{\sun}$, semimajor axis of $20\,\mbox{AU}$, and $e=0$.
The results of $10^4$ runs are shown in Figure \ref{fig:cfperi}, in which
we plot the cumulative fraction of binaries as a function of $r_p/a$, where
$r_p$ is the pericentre distance of the merger binary, and $a$ is the initial
binary semimajor axis.  The agreement between {\em scatter3\/} (solid line) and 
{\em scatter\/} (dashed line) is good, with both yielding merger binaries with $r_p$ strongly
concentrated between $0.15\,\mbox{AU}$ and $0.3\,\mbox{AU}$.

\section{Systematic Study of the Collision Cross Section}

To better understand the behavior of the collision cross section, we have 
systematically studied its dependence on several physically
relevant parameters.  The understanding gained
will allow us to reduce the dimensionality of parameter space that must be
sampled when we later consider MS-star binaries with physically motivated
parameters.

\subsection{Dependence on velocity at infinity}

The dimensionless collision cross section ($\sigma/(\pi a^2)$ for binary--single, 
$\sigma/(\pi (a_0+a_1)^2)$ for binary--binary) 
as a function of the relative velocity at infinity, $v_\infty/v_c$, is shown
in Figure \ref{fig:v_sigma}, for both binary--single interactions (left) and 
binary--binary interactions (right), for several different values of the expansion
parameter, $f_{\rm exp}$.
Circles represent outcomes with one or more collisions (two or more stars collide); 
triangles, two or more (three or more stars collide); and
squares, three (four stars collide).  Red represents runs with $f_{\rm exp}=1$; orange,
$f_{\rm exp}=2$; green, $f_{\rm exp}=5$; and blue, $f_{\rm exp}=10$.
In both experiments (binary--single and binary--binary), each star had mass
$M_{\sun}$ and radius $R_{\sun}$, and each binary had semimajor axis 
$a=1\,\mbox{AU}$ and eccentricity $e=0$.  The cross section decreases
sharply at $v_\infty/v_c=1$, above which resonant scattering is forbidden, and appears
to approach a constant value, consistent with being purely geometrical.
In the resonant scattering regime, below $v_\infty/v_c=1$, the collision 
cross section follows the form $1/v_\infty^2$, implying that gravitational focusing 
is dominant.  The $n_{\rm coll} \geq 1$ cross section in the resonant scattering
regime is quite high, with $\sigma (v_\infty/v_c)^2/(\pi a^2)\approx 1$ for 
binary--single and $\sigma (v_\infty/v_c)^2/(\pi (a_0+a_1)^2) \approx 0.8$ for binary--binary.

The $n_{\rm coll} \geq 2$ cross section in the binary--single case is about two to
three orders of magnitude below that for $n_{\rm coll} \geq 1$, depending on $f_{\rm exp}$.  
However, in the binary--binary case, the $n_{\rm coll} \geq 2$ cross section is only 
down by a factor of a few to 10.  The reason for the difference is that in the binary--single
case, after one collision occurs, there are only two stars left.  The two remaining stars
will either be bound in a binary, or unbound to each other in a hyperbolic orbit.  In 
the case of a bound orbit, the two stars are guaranteed to make at least one pericentre
passage, and if the merger product in the binary is large enough, a collision will occur.
In the case of an unbound orbit, the likelihood of a pericentre passage is decreased.
In either case, it is clear that with only two stars remaining, the complex resonant
behavior observed in three- and four-body interactions that leads to close approaches 
will not occur.  

There is a large spread in the $n_{\rm coll} \geq 3$ cross section in binary--binary scattering.
This is because it is likely for collision products to suffer subsequent collisions
given their increased size, implying that the $n_{\rm coll} \geq 3$ cross section
should vary as $f_{\rm exp}^2$.  The $n_{\rm coll} \geq 3$ cross section varies from a factor
of a few to two orders of magnitude below that for $n_{\rm coll} \geq 2$.

Finally, we note that the spread in the $n_{\rm coll} \geq 2$ binary--binary cross
section is a factor of about four, essentially independent of $v_\infty$ for $v_\infty/v_c \la 1$, 
as $f_{\rm exp}$ varies over an order of magnitude.
The cross section is therefore not a particularly sensitive function of the unknown expansion
parameter $f_{\rm exp}$, and, if it is valid to parametrize the size of collision
products in this simplified manner, implies that our results for the properties of
merger populations are relatively robust.

\subsection{Dependence on the ratio of stellar radius to binary semimajor axis}\label{sec:r_sigma}

The collision cross section varies as $1/v_\infty^2$ for $v_\infty/v_c<1$,
the regime relevant to interactions involving hard binaries in the cores of 
globular clusters.  Therefore, we can choose a single value for $v_\infty$ when
exploring the dependence of the collision cross section on other physically
relevant parameters, thereby reducing the dimensionality of parameter space
that must be sampled.  For the remainder of this section, we set $v_\infty/v_c=0.1$,
which corresponds to typical binary--single and binary--binary interactions involving
hard binaries in a globular cluster core, with $v_\infty=10\,\mbox{km/s}$, 
stars of mass $M_{\sun}$, radius $R_{\sun}$, and binaries with 
$a=0.1\,\mbox{AU}$.

Figure \ref{fig:r_sigma} shows the normalized, dimensionless collision cross section,
$\sigma (v_\infty/v_c)^2/(\pi a^2)$ for binary--single scattering (left),
$\sigma (v_\infty/v_c)^2/(\pi (a_0+a_1)^2)$ for binary--binary scattering (right),
as a function of the ratio of stellar
radius to binary semimajor axis, $R/a$, for different values of the expansion 
parameter, $f_{\rm exp}$.  Circles represent outcomes with one or more collisions; 
triangles, two or more; and squares, three or more.  Red represents runs with 
$f_{\rm exp}=1$; orange, $f_{\rm exp}=2$; green, $f_{\rm exp}=5$; and blue, $f_{\rm exp}=10$.
In both experiments (binary--single and binary--binary), each star had mass $M_{\sun}$ and radius $R$,
each binary had semimajor axis $a=1\,\mbox{AU}$ and eccentricity $e=0$, and the relative velocity
at infinity was set to $v_\infty/v_c=0.1$.  Calculations were performed down
to $R/a=10^{-9}$ -- which corresponds to the extreme case of binaries with semimajor
axis $10\,\mbox{AU}$ composed of black holes of mass $M_{\sun}$ -- but no collisions 
were found below $R/a\approx 10^{-6}$.  For $n_{\rm coll} \geq 1$, the calculation
corresponds to the simpler task of recording minimum close approach distances,
as can be seen by comparing the binary--binary panel (right) to Figure
\ref{fig:baconcomp}.  The $n_{\rm coll} \geq 2$ and $n_{\rm coll} \geq 3$ collision 
cross sections decrease more sharply than the $n_{\rm coll} \geq 1$ cross section
as $R/a$ decreases.  

It is clear that multiple collisions are unlikely for
$R/a \la 0.001$, which corresponds roughly to stars of radius $R_{\sun}$ in binaries
with semimajor axis $1\,\mbox{AU}$.  We therefore expect that multiple collisions
in binary interactions are relevant only for MS stars in binaries tighter
than $\sim 1\,\mbox{AU}$, white dwarfs in binaries tighter than $\sim 1\,R_{\sun}$,
and neutron stars in binaries tighter than $\sim 10^{4}\,$km.  We caution that
relativistic effects may need to be included when considering close approaches
of neutron stars.  However, the limits quoted should serve as a rough guide.

We have held the stellar masses fixed at $M_{\sun}$, while varying their radii
over a large range.  For MS stars, it is more realistic to adopt a
reasonable mass-radius relationship, which we do in \S~\ref{sec:realistic} for 
several sets of masses.

\subsection{Dependence on mass ratio}

In binary interactions involving stars of different masses, there is a strong
tendency for the lightest star(s) to be ejected quickly
\citep[see, e.g.,][]{2003gmbp.book.....H}.  One would expect, then, that 
resonant behavior, and the likelihood of collisions, would be decreased when one or
more of the stars involved are light.  To test this prediction, we have calculated
the collision cross section during binary--single and binary--binary scattering 
for a range of mass ratios.
In both experiments, each binary had one star with mass $M_{\sun}$ and the 
other with mass $q M_{\sun}$.  For the binary--single case, the incoming single 
star had mass $M_{\sun}$.  Each star had radius $R_{\sun}$, each binary had 
semimajor axis $a=1\,\mbox{AU}$ and eccentricity $e=0$, and the relative velocity at 
infinity was set to $v_\infty/v_c=0.1$.  We normalize the cross section, as usual, by
multiplying by $(v_\infty/v_c)^2$, and, in doing so, inadvertently introduce a dependence
on the mass ratio, $q$, in $v_c^2$.  To remove it, we also multiply by a function of $q$ alone
that has the same dependence on $q$ as $v_c^2$, from eqs.~(\ref{eq:vcbs}) and 
(\ref{eq:vcbb}), and is normalized to 1 at $q=1$.  For binary--single interactions this function
is $2q(2+q)/(3(1+q))$; for binary--binary interactions it is $2q/(1+q)$.  The
collision cross sections are shown in Figure \ref{fig:q_sigma} for binary--single
(left) and binary--binary (right), as a function of $q$, for different values
of the expansion parameter, $f_{\rm exp}$.  Circles represent outcomes with one or 
more collisions; triangles, two or more; and squares, three or more.  Red represents 
runs with $f_{\rm exp}=1$; orange, $f_{\rm exp}=2$; green, $f_{\rm exp}=5$; and 
blue, $f_{\rm exp}=10$.  As expected, the collision cross section is smaller for 
$q<1$.  However, it decreases quickly, and for $q \la 0.1$ becomes approximately
constant, implying that the test particle limit has been reached.  What is most
striking is that the collision cross section is decreased by no more than a factor
of a few for small $q$, despite the tendency for lighter stars to be ejected
quickly.
It should be noted that in this experiment we have kept the radii of all stars fixed 
at $R_{\sun}$.

\section{Results for typical binaries}\label{sec:realistic}

We now turn from a slicing of parameter space to a discrete sampling, by 
considering binaries with sets of parameters typical of those found in the cores
of globular clusters.  We first present results for binary--single interactions,
and then binary--binary.

\subsection{Binary--single scattering experiments}

We consider only MS stars with masses $0.5\,M_{\sun}$, 
$1.0\,M_{\sun}$, or $1.2\,M_{\sun}$.  We adopt the mass-radius relationship 
$R = R_{\sun} (M/M_{\sun})$, which is a reasonable approximation
for MS stars of mass $\sim 1\,M_{\sun}$.  We study five
different mass combinations, labelled A through E, with a range of
semimajor axes, $0.05\,\mbox{AU}\leq a \leq 3.0\,\mbox{AU}$, for each.
In all cases we use $v_{\infty}=10\,\mbox{km/s}$.  This choice of parameters
covers a range of binary binding energies from $\sim 1\,kT$ (the 
hard-soft boundary) in a typical globular cluster core, to $\sim 10^2\,kT$, 
corresponding to a close binary ($a\sim 10 R_{\sun}$).  The thermal energy $kT$ is defined
by the relation $\frac{1}{2}kT = \frac{1}{2}\langle m \rangle \sigma^2$,
where $\langle m \rangle$ is the average stellar mass, and $\sigma$ is
the one-dimensional velocity dispersion.
The details of each run are presented in Table~\ref{3body},
including run name; the number of scattering interactions performed, $N$;
the masses of the binary members, $m_{00}$ and $m_{01}$; the mass of the intruder,
$m_1$; the binary semimajor axis, $a$; and the $n_{\rm coll} \geq 1$ cross section.

In order to study the dependence of the collision cross section on the
expansion parameter, $f_{\rm exp}$, without performing calculations for each value
of $f_{\rm exp}$ considered, we have adopted an approach that allows us to calculate
multiple collision cross sections for any value of $f_{\rm exp}$ based on the results
of calculations for one value of $f_{\rm exp}$.  We set $f_{\rm exp}=1$, and consider the 
properties of merger binaries formed.  A binary containing a merger product will
be a triple-star merger if the pericentre of the binary, $r_p$, is approximately less
than the radius of the collision product, $R_{\rm cp}=f_{\rm exp}(R_1+R_2)$, where
$R_1$ and $R_2$ are the radii of the two stars that merged to form the collision 
product.  First we calculate
$N_{\rm coll}$, the total number of outcomes that resulted in either merger binaries
or triple mergers with $f_{\rm exp}=1$.  We then calculate $N_{\rm 3coll}$, the number
of triple mergers, for a different value of $f_{\rm exp}$, as the number
of triple mergers for $f_{\rm exp}=1$, plus the number of merger binaries
with $r_p<R_{\rm cp}$.  Defining $f_T=N_{\rm 3coll}/N_{\rm coll}$, the triple-star
merger ($n_{\rm coll} \geq 2$) cross section for $f_{\rm exp}$ is simply 
$\sigma_T(f_{\rm exp})=f_T \sigma_{\rm coll}(f_{\rm exp}=1)$.

Some remarks about this approach are in order.  We ignore merger
escapes, and argue that an outcome labelled as a merger
escape is unlikely to become a triple merger even if the first merger
product expands.  Before it escapes, the third star can
approach the expanded merger at most once, and, if it does, it is
likely to have a sufficiently high speed at close approach to fully
traverse the tenuous envelope of the expanded merger product. On the contrary, in a merger
binary, even if the third star initially has a high pericentric speed,
it will eventually be captured through gradual energy loss after
repeated traversal. Of course, an escaping third star may lose
sufficient energy after traversal so that the entire system becomes
bound, and eventually be captured. A more precise treatment would
be to run calculations for each value of $f_{\rm exp}$, but, as mentioned
above, we are adopting the simpler, less computationally expensive approach here.
When two MS stars collide and their merger
product expands, the resulting object does not possess a well-defined
boundary and, in general, is not spherically symmetric; $f_{\rm exp}$ is 
thus an effective, averaged quantity, which serves well enough the 
purpose of our first study.

\subsubsection{Collision cross sections}

The $n_{\rm coll} \geq 1$ cross sections are listed
in the last column of Table \ref{3body}.  The cross sections from runs 
A, B, and~C are also shown as a function of the initial binary semimajor
axis, $a$, in Figure~\ref{fig:abc}.  In the range of MS masses
of interest for globular clusters, the collision
cross sections show only a weak dependence on masses, slightly more pronounced
at small $a$.  The cross section increases from case A to C 
as the mass ratios of the stars decrease, due to the dependence of the normalized
cross section on $v_\infty/v_c$ and hence on the mass ratio.
For a hard binary with $a \sim 1\,\mbox{AU}$, the 
normalized collision cross section is comparable to the geometric
cross section of the initial binary 
(i.e., $\sigma_{\rm coll} (v_\infty/v_c)^2 \sim \pi a^2$). This is because
most strong interactions are resonant, and most resonances lead to at least 
one collision.  For $a \la 0.1\,\mbox{AU}$, the collision cross section can 
be up to an order of magnitude greater than the geometric cross section. 
Indeed, for very small values of $a$, even a small perturbation of a 
highly eccentric orbit by a distant encounter can induce a binary merger. 
About 20 to 35\% of the initial binaries with $a=0.05\,\mbox{AU}$ in 
Table \ref{3body} have pericentre distances less than $3\,R_{\sun}$. 
Our results for these very tight binaries are therefore somewhat
artificial, since in reality tidal circularization effects are likely
to modify the distribution of initial eccentricities, and our simple
assumption of a thermal initial distribution is no longer justified.

\subsubsection{Properties of the merger binaries} 

Of particular interest are binary--single interactions that result
in binaries containing merger products.  The distributions of their
properties are relevant to observations of BSs in the cores
of globular clusters.  Figures~\ref{mb1} and~\ref{mb2} show
the orbital parameters of the merger binaries produced in the two
representative runs A300, for a wide initial binary, and B005, for a very
tight initial binary.  The envelope of the distribution follows curves
of constant angular momentum, consistent with angular momentum conservation
during the interaction.  The total angular momentum of the system is the sum of
the initial internal angular momentum of the binary and the initial angular momentum
of the binary--single hyperbolic orbit, added vectorially.  The spread in angular
momentum spanned by the distributions is due to averaging over 
the relative orientation of the two separate angular momenta, the range of 
initial eccentricities of the binary, and range of impact parameters used.  
Curves of constant angular momentum are plotted in Figure~\ref{mb1}, 
for the values $J/J_0=0.2$, 0.5, 1.0, and 2.0, where 
$J_0=\mu b v_\infty + \mu_b [GM_b a(1-e^2)]^{1/2}$ is the angular
momentum of the system such that the pericentre distance of the initial hyperbolic
orbit is $1.0\,\mbox{AU}$ (i.e., $b=r_p(1+2GM/r_p v_\infty^2)^{1/2}$ with $r_p=1.0\,\mbox{AU}$).
(Here $\mu$ and $M$ are the reduced and total mass of the binary--single system, and
$\mu_b$ and $M_b$ are the reduced and total mass of the binary.)
The vertical dashed line in Figures 
\ref{mb1} and \ref{mb2} is the hard-soft boundary with respect to field stars 
of mass $1\,M_{\sun}$ with one-dimensional velocity dispersion
$10\,\mbox{km/s}$.  Histograms of final semi-major axes and eccentricities 
are shown in Figures \ref{hista} and \ref{histe}.  The dotted lines in Figure
\ref{histe} are properly normalized thermal eccentricity distributions.

Typically, more than 90\% of the merger binaries have final semimajor axis,
$a^\prime$, larger than initial, $a$.  On average, $a^\prime/a \approx 5$. 
While most remain hard binaries, a small fraction become soft, with a few having $a^\prime$ as
large as $\sim 100-1000\,\mbox{AU}$. This softening comes from the
somewhat counter-intuitive result that collisions produce, on
average, an {\em increase\/} in the {\em orbital energy\/} 
of the system (while the total energy, including the binding energy
of the collision product, is of course conserved on a dynamical
time-scale, i.e., until some of the internal energy released through
shocks can be radiated away by the fluid).
To illustrate this, consider a trivial example in which two identical
stars of mass $m$ are released from rest at some distance $r$ and
collide head-on, forming a stationary merger product at the centre of 
mass.  The orbital energy of the system increased by $Gm^2/r$ in the
process. More relevant to our results, but still somewhat artificial,
consider an initial binary with a very high eccentricity, so that
the two members almost collide at pericentre.  A small perturbation
through a distant encounter can induce a merger of the binary (implying
that its orbital binding energy disappears), while only weakly affecting
the orbit of the perturber.

The eccentricity distributions of merger binaries always remain close to
thermal, although a slight excess of highly eccentric orbits is seen
for wider initial separations (compare run A300, with $a=3\,\mbox{AU}$, and
B005, with $a=0.05\,\mbox{AU}$, in Figure \ref{histe}). 
The average value of $e^\prime$ for runs A300 and B005 is 
$0.77$ and $0.68$, respectively, while that of a thermal distribution is $2/3$.
It is interesting to note that other calculations of small-$N$ systems have yielded
binaries with an excess of high eccentricity systems in a nearly thermal distribution
\citep{2000ApJ...528L..17P}.

\subsubsection{Three-star mergers} 

Three-star mergers happen primarily when the pericentre distance of a merger
binary is approximately smaller than the radius of the merger remnant.
Cumulative distributions of pericentre distances from all A runs
are shown in Figure \ref{cumfrac}.  For radii of first collision
products in the range $\sim5$--$10\,R_{\sun}$ ($f_{\rm exp}\sim 2.5$--$5$), 
we find triple collision
fractions anywhere from a few percent up to 50\%, depending strongly
on the initial binary semi-major axis $a$. Clearly, triple collisions
occur often, particularly during encounters with very hard binaries.
If we consider the later expansion of the collision product on the
giant branch (with radius up to $\ga 100\,\mbox{AU}$), a triple collision
becomes almost inevitable, except for only the widest initial binaries.

Denoting the value of $R_{\rm cp}$ at which $f_T=f$ by $R_f$,
we determine the critical radii of merger products corresponding to a given
triple collision fraction, $R_{0.05}$, $R_{0.1}$, $R_{0.5}$, $R_{0.9}$ and 
$R_{0.95}$, using simple linear interpolation.
These are plotted as a function of the initial binary separation $a$
in Figure \ref{rf}. The error bars in Figure \ref{rf} are estimated by
dividing the uncertainty in $f_T$ by the slope of the $f_T$ vs $R_{\rm
cp}$ curve at $R_{\rm cp}=R_f$, i.e.,
\begin{equation}
\Delta R_f \simeq \frac{\sqrt{f/N_{\rm coll}}}{(df/dR_{\rm cp})_{R_f}} \, .
\end{equation}
We see that all the lines in Figure \ref{rf} are nearly parallel and 
with a slope close to unity.  The same holds true for mass combinations B
through E as well.  Thus we have approximately $R_f\propto a$
and the relationship can be specified by a single quantity $R_f/a$
for each value of $f_T$. These have been
estimated using a least-squares fit with weights inversely proportional
to the size of the error bars. Since hydrodynamic calculations have
shown that $R_{\rm cp}$ is unlikely to be larger than $\sim30$ times the
original stellar radius, according to Figure \ref{rf}, the most
relevant range corresponds to $f_T\la 0.5$
(although the full range up to $f_T\approx 1$ will be relevant if 
the later expansion of the merger product on the red giant branch 
is considered). In Figure \ref{raf} we plot $R_f/a$
as a function of $f_T$.  It is clear that $R_f/a$
is directly proportional to $f_T$ over the range of
interest. For run A (equal mass case), the proportionality constant 
is $1.61\pm 0.01$. Consequently, the relation between $R_{\rm cp}$, $f_T$
and $a$ for this particular mass combination may be written
\begin{equation}\label{eqn:raf111}
R_{\rm cp} \approx 1.6 a f_T \, ,
\end{equation}
where $0.05\,\mbox{AU}\leq a\leq 3.0\,\mbox{AU}$.
Turning to different mass combinations we find results similar to eq.~(\ref{eqn:raf111}), 
and so can write
\begin{equation}\label{eqn:raf}
R_{\rm cp} = C a f_T \, ,
\end{equation}
where $C$ depends only on the stellar masses. Table \ref{coef}
shows $C$ for the five mass combinations we have explored.

\subsection{Binary--binary scattering experiments}

For the sake of convenience, we use the abbreviations listed in 
Table \ref{tab:bboutcomes} to refer to certain binary--binary outcomes.  In
the abbreviated form, the letters S, D, T, and Q denote a single star, 
double-star merger, triple-star merger, and quadruple-star merger, respectively,
and we have also chosen to use parentheses instead of square brackets.
Each run we do involves MS stars of either $0.5\,M_{\sun}$ or
$1.0\,M_{\sun}$ and binary semimajor axes of either $1.0\,\mbox{AU}$ or 
$0.1\,\mbox{AU}$.  In all runs we set $v_\infty=10\,\mbox{km/s}$, as in 
the binary--single case.  The properties of each run are listed in 
Table~\ref{4body}, including the mass of each star, $m_{ij}$, the
semimajor axis of each binary, $a$, and the normalized cross sections
for strong interactions and at least one collision to occur.

To study the dependence of the outcomes on the expansion parameter,
$f_{\rm exp}$, we have performed separate calculations for each value of 
$f_{\rm exp}$ considered.  For binary--binary interactions, the dynamics 
do not reduce to the trivial analytical case of two-body motion after one
collision has occurred, and so it is not possible to use the simple approach
of tracking pericentre distances in merger binaries as we did for the 
binary--single case.  It should be noted, however, that we apply the simple expansion
factor prescription for the radius of a merger product, 
$R_{\rm merger}=f_{\rm exp}(R_1+R_2)$, where $R_1$ and $R_2$ are the radii of the merging
stars, to every merger, regardless of whether the merging stars are unperturbed
MS stars or merger products themselves.  The simplicity of this
prescription allows us to study the dependence of our results on only one
parameter, $f_{\rm exp}$, which can thus be considered an {\em effective\/} 
expansion parameter, averaged over all types of mergers.  
A more realistic approach that adopts separate expansion
parameters for different types of mergers is feasible, but beyond the scope
of this study.

\subsubsection{Collision cross sections}

The normalized cross sections for strong interactions, $\sigma_{\rm strong}$, and 
for at least one collision, $\sigma_{\rm coll}$, for our binary--binary runs are 
listed in the last two columns of Table~\ref{4body}.  A strong interaction is defined to be
one in which the final configuration is different from the initial
configuration (i.e., anything but preservation), or a preservation resulting from a
resonant encounter.  The test for a resonant encounter is that of \citet{1983ApJ...268..319H}, 
wherein the mean square distance between pairs of stars is checked for multiple minima.

Comparing the results from run~I ($a_0=a_1=1\,\mbox{AU}$) with run~II ($a_0=a_1=0.1\,\mbox{AU}$), 
we see that $\sigma_{\rm coll}$ is a larger fraction of $\sigma_{\rm strong}$ for 
run~II, consistent with our findings in \S~\ref{sec:r_sigma} for small $R/a$.  Comparing run~II with
run~IV, we see that introducing a non-unity mass ratio does not seem to affect
$\sigma_{\rm strong}$, but slightly lowers $\sigma_{\rm coll}$.  By calculating the
branching ratio for outcome X involving collisions -- defined as $f_X=N_X/N_{\rm coll}$
where $N_{\rm coll}$ is the number of outcomes that result in collisions and 
$N_X$ is the number of those that result in outcome X -- the value 
of $\sigma_{\rm coll}$ for a particular run can be used to calculate the cross 
section for outcome X, according to the simple relation $\sigma_X = f_X \sigma_{\rm coll}$.

\subsubsection{Properties of merger products} 

In Figures \ref{br1} and \ref{br2}, we show the branching ratios for several
outcomes as a function of $f_{\rm exp}$.  That is, we plot the fraction of 
outcomes involving at least one collision that result in various configurations
containing double-star, triple-star, and quadruple-star mergers.  Figure~\ref{br1}
shows results from run~I ($a_0=a_1=1\,\mbox{AU}$) and Figure~\ref{br2} shows
results from run~II ($a_0=a_1=0.1\,\mbox{AU}$).  The upper left panel in each shows
the branching ratios for outcomes of two unbound double-star mergers, labelled DD, 
and two double-star mergers in a binary, labelled (DD); the upper right, a quadruple-star merger,
labelled Q; the
lower right, a triple-star merger bound to the remaining single star, labelled (TS); 
and the lower left, the combined branching ratio for any outcome involving a merger of three
or more stars, labelled T/Q.

From Figure \ref{br1}, we see that, even for encounters involving wider binaries,
the branching ratio for more than two stars to merge is significant -- as high as $\sim 5\%$.
When one considers tighter binaries, as in Figure~\ref{br2}, the branching ratio
increases to $\sim 40\%$.  The dependence on initial semimajor axis
is as expected -- all branching ratios for mergers are increased in run~II over
run~I.  The dependence on $f_{\rm exp}$ is also as expected.  As the expansion
factor is increased, more multiple mergers occur, leading to an increase in the branching
ratios for triple-star and quadruple-star mergers, and a decrease in those for
double-star mergers.

The distributions of orbital parameters for all four types of binaries, (DS)S, D(SS), (DD), and (TS)
are plotted in Figures \ref{ae1_5} and \ref{ae2_5} for runs I and II, with $f_{\rm exp}=5$.
From these figures, we see that (DS) binaries form with semimajor axes comparable to, 
and only slightly greater than, the semimajor axes of their progenitor binaries (except for the
case (DS)S, where $a^\prime$ can be significantly larger than $a$ for large $e^\prime$), and with
an eccentricity distribution that does not appear to be inconsistent with thermal.
The data are more sparse for the (DD) and (TS) cases, but their orbital parameters appear
to be comparable to those of (DS) binaries.

The three outcomes TS, (TS), and Q (labelled T/Q collectively in Figures \ref{br1} and 
\ref{br2}), are responsible for the production of BSs of mass $>2M_{\sun}$
in our runs.  The branching ratio for T/Q appears to increase almost linearly with 
$f_{\rm exp}$ in the range considered, for all runs performed.  Linear fits for the 
branching ratio of T/Q as a function of $f_{\rm exp}$ (obtained by least squares fitting)
are provided in Table \ref{fit}.

\section{Summary, conclusions, and future directions}

We have performed several sets of binary--single and binary--binary scattering experiments,
and studied the likelihood of (multiple) collisions.  We have presented collision cross
sections, branching ratios, and sample distributions of the parameters of outcome
products.  Results reported in this paper, particularly cross sections, may be employed 
in both analytical and numerical calculations.

In the gravitational focusing regime, relevant to hard binaries in globular cluster cores, 
the likelihood of collisions during binary interactions is quite high.  
For solar mass main-sequence (MS) stars in $1\,\mbox{AU}$ binaries, the normalized cross section for at least
one collision to occur during a binary--single or binary--binary interaction 
($n_{\rm coll} \geq 1$) is essentially unity, with 
$\sigma (v_\infty/v_c)^2 / (\pi a^2) \sim 1$ for binary--single and 
$\sigma (v_\infty/v_c)^2 / (\pi (a_0+a_1)^2) \sim 1$ for binary--binary.
The collision cross section depends strongly on the ratio of stellar radius to binary
semimajor axis, but is reasonably high even for MS stars of approximately solar mass in orbits of
$\sim 1\,\mbox{AU}$.  Perhaps counter to intuition, the
collision cross section is not particularly sensitive to binary mass ratio, dropping by
only a factor of a few in the test-particle limit when the stellar
radii are kept fixed.  We also found that the multiple collision
($n_{\rm coll} \geq 2$) cross section is quite high, only a factor of $\sim 10$ lower
than the $n_{\rm coll} \geq 1$ cross section for binary--binary interactions.  It is 
also not a particularly sensitive function of the expansion parameter, $f_{\rm exp}$, 
varying by a factor of a few as $f_{\rm exp}$ is varied by an order of magnitude.
This implies that studies using this one-parameter model for the radius of a collision
product are reasonably robust in spite of the large uncertainties in the physics.
For typical binaries in globular cluster cores,
we have shown that collisions of more than two stars during binary--single
and binary--binary interactions are likely, with branching ratios for triple-star
mergers of $\sim 5\%$ for binary--single and $\sim 10\%$ for binary--binary.

We have introduced \Fewbody, a new numerical toolkit for simulating small-$N$
gravitational dynamics that is particularly suited to performing scattering
interactions.  We have shown that it produces results in good agreement with several 
previous numerical studies of binary--single and binary--binary scattering, as well
as with the \Starlab\ software suite.  Instead of using cross sections and simple recipes
for binary interactions in globular cluster evolution codes, one may use \Fewbody\ to perform
them directly.  We have adopted this approach with our Monte Carlo globular cluster
evolution code \citep{2003ApJ...593..772F}.

It is clear from our results that collisions of more than two stars during binary
interactions are a viable pathway for creating blue stragglers (BSs) with masses more
than twice the MS turnoff mass, such as those observed in NGC~6397 \citep{2000AAS...196.4106S}.
These massive BSs may also be formed via recycling -- in other words, a binary containing a BS
may be formed via a binary interaction, and the BS may later merge with
another star in a subsequent binary interaction, creating a more massive
BS.  We are in the process of creating a more detailed model, based on
a Monte Carlo binary population study, which incorporates both channels
to study the formation of massive BSs.  Such studies are needed to help
interpret current BS observations \citep[see][]{1999ApJ...513..428S,2000ApJ...535..298S}, 
and the large databases of BS properties, including many new spectroscopic
mass measurements, that will soon be available (M.~Shara, private communication).

The expansion of merger products has been treated here in a simplified manner, using a single
expansion parameter, $f_{\rm exp}$.  As observations of BSs become more detailed and more numerous,
including details of their internal properties, it becomes necessary to treat collisions
in a more accurate way.  Full smoothed particle hydrodynamics (SPH) calculations are quite
computationally prohibitive, taking up to several hours to perform a single merger.
However, there are faster, approximate approaches that capture the essential physics of
the hydrodynamic merger process.  One such approach is the fluid-sorting algorithm, which utilises
the property that the fluid in merger products must rearrange itself according to specific
entropy \citep{1995ApJ...445L.117L,2002ApJ...568..939L}.
The {\em Make Me A Star\/} (MMAS) software developed by Lombardi and collaborators implements this 
procedure, and is freely available on the web 
\citep{1996ApJ...468..797L,2002ApJ...568..939L,2003MNRAS.345..762L}.  
We have begun to replace the simple merger module in
\Fewbody\ with a call to MMAS.  The result should be much more
accurate predictions for the properties of (multiple) merger products.

\section*{Acknowledgments}
We thank Steve McMillan, Steinn Sigurdsson, and Saul Rappaport
for many helpful discussions.  This work was supported by NSF Grant AST-0206276 and
NASA ATP Grant NAG5-12044.
J.M.F.\ would like to acknowledge the hospitality
of the Theoretical Astrophysics Group at Northwestern University.
P.C.\ acknowledges partial support from the UROP 
Program at MIT.  S.P.Z.\ was supported by
the Royal Netherlands Academy of Sciences (KNAW), the Dutch
Organization of Science (NWO), the Netherlands Research School for Astronomy (NOVA), 
and Hubble Fellowship grant HF-01112.01-98A awarded by
the Space Telescope Science Institute, which is operated by the
Association of Universities for Research in Astronomy, Inc., for NASA
under contract NAS\, 5-26555. 

\bibliographystyle{mn}
\bibliography{mn-jour,main}

\clearpage

\begin{table}
\begin{center}
\caption{\label{tab:bsoutcomes}Possible outcomes of binary--single star encounters, ordered by the number of 
  collisions, $n_{\rm coll}$.  
  Brackets enclose two objects which are bound to each other, while colons represent 
  physical collisions.  For simplicity, we have only listed the outcomes that
  would result from indistinguishable stars.}
\begin{tabular}{llll}
\hline
$n_{\rm coll}$& symbol &description\\
\hline 
0 &\null[$\bullet$ $\bullet$] $\bullet$ &preservation or exchange\\
0 &\null$\bullet$ $\bullet$ $\bullet$ &ionization\\
0 &\null[[$\bullet$ $\bullet$] $\bullet$] &stable hierarchical triple\\
\hline
1 &\null[$\bullet$:$\bullet$ $\bullet$] &binary containing a two-star merger\\
1 &\null$\bullet$:$\bullet$ $\bullet$ &two-star merger and single star\\
\hline
2 &\null$\bullet$:$\bullet$:$\bullet$ &three-star merger\\
\hline
\end{tabular}
\end{center}
\end{table}

\begin{table}
\begin{center}
\caption{\label{tab:bboutcomes}Possible outcomes of binary--binary star encounters, ordered by the number 
  of collisions, $n_{\rm coll}$.  Brackets enclose two objects which are bound to each other, while colons represent 
  physical collisions.  For simplicity, we have only listed the outcomes that
  would result from indistinguishable stars.  Listed in the third column are the abbreviations used in
  the paper to refer to various outcomes.}
\begin{tabular}{llll}
\hline
$n_{\rm coll}$& symbol &abbreviation &description\\
\hline 
0 &\null[$\bullet$ $\bullet$] [$\bullet$ $\bullet$] & &preservation or exchange\\
0 &\null[$\bullet$ $\bullet$] $\bullet$ $\bullet$ & &single ionization\\
0 &\null$\bullet$ $\bullet$ $\bullet$ $\bullet$ & &full ionization\\
0 &\null[[$\bullet$ $\bullet$] $\bullet$] $\bullet$ & &stable hierarchical triple and single star\\
0 &\null[[[$\bullet$ $\bullet$] $\bullet$] $\bullet$] & &stable hierarchical quadruple\\
0 &\null[[$\bullet$ $\bullet$] [$\bullet$ $\bullet$]] & &stable quadruple composed of two binaries\\
\hline 
1 &\null[$\bullet$ $\bullet$] $\bullet$:$\bullet$ &(SS)D &binary and two-star merger\\
1 &\null[$\bullet$:$\bullet$ $\bullet$] $\bullet$ &(DS)S &single star and binary containing two-star merger\\
1 &\null$\bullet$:$\bullet$ $\bullet$ $\bullet$ &DSS &two-star merger and two single stars\\
1 &\null[[$\bullet$:$\bullet$ $\bullet$] $\bullet$] &((DS)S) &stable hierarchical triple with two-star merger in inner binary\\
1 &\null[[$\bullet$ $\bullet$]  $\bullet$:$\bullet$] &((SS)D) &stable hierarchical triple with two-star merger in outer binary\\
\hline 
2 &\null[$\bullet$:$\bullet$ $\bullet$:$\bullet$] &(DD) &binary composed of two two-star mergers\\
2 &\null[$\bullet$:$\bullet$:$\bullet$ $\bullet$] &(TS) &binary containing a three-star merger\\
2 &\null$\bullet$:$\bullet$ $\bullet$:$\bullet$ &DD &two two-star mergers\\
2 &\null$\bullet$:$\bullet$:$\bullet$ $\bullet$ &TS &three-star merger and single star\\
\hline 
3 &\null$\bullet$:$\bullet$:$\bullet$:$\bullet$ &Q &four-star merger\\
\hline
\end{tabular}
\end{center}
\end{table}

\clearpage
\begin{figure}
  \begin{center}
    \pstree[levelsep=40pt]{\Tcircle{0}}{}
    \quad
    \pstree[levelsep=40pt]{\Tcircle{\phantom{0}}}{
      \pstree[levelsep=40pt]{\Tcircle{\phantom{0}}}{\Tcircle{1}\Tcircle{2}}
      \Tcircle{3}
    }
    \quad
    $\cdots$
    \quad
    \pstree[levelsep=40pt]{\Tcircle{$N$}}{}
    \caption{\label{fig:secondpasstree}Schematic representation of the binary-tree algorithm used
      by \Fewbody.  A circle containing a number represents a star, while an empty circle represents
      a general parent node.  The set of binary trees is shown after the second stage of processing,
      with the 1-2 centre of mass and star 3 replaced by their parent node.  For hierarchy
      isolation, this replacement is quite rare, as it would require that the triple be not only
      dynamically stable, but also sufficiently hierarchical that its evolution could be treated
      analytically.}
  \end{center}
\end{figure}

\begin{figure}
  \begin{center}
    \includegraphics[width=0.48\textwidth]{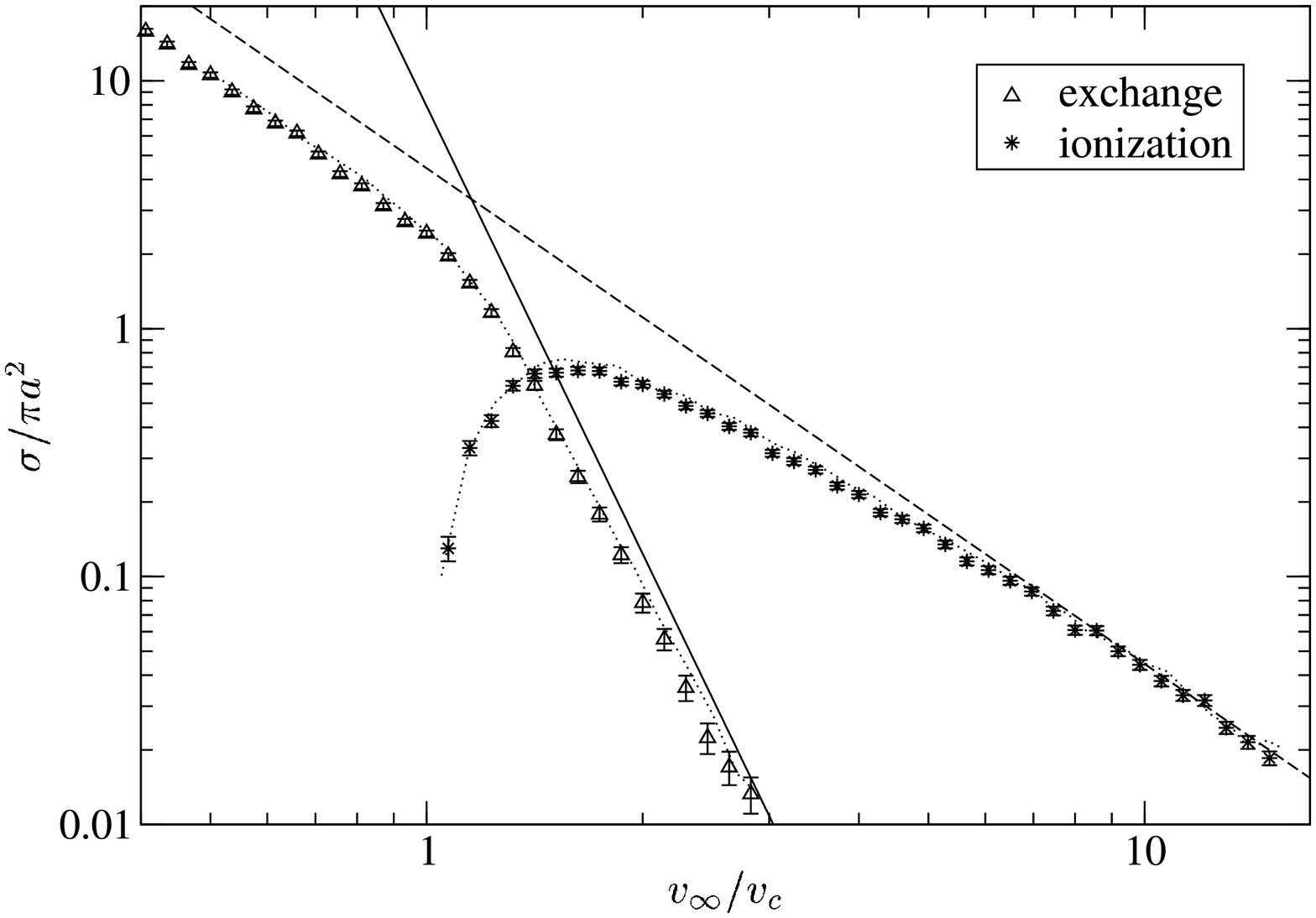}
    \caption{Comparison of \Fewbody\ with Figure 5 of \citet{1983ApJ...268..319H}: 
      total cross sections for
      binary--single scattering for the equal-mass, zero eccentricity, point-particle case.  A total
      of $8\times 10^5$ scattering experiments were used to create this figure.  The dotted
      lines represent the data from \citet{1983ApJ...268..319H}, while the straight
      solid and dashed lines are the theoretically predicted cross sections for ionization and 
      exchange from the same paper.  Data points are from \Fewbody.  The agreement between
      the two is excellent.\label{fig:hutcomp}}
  \end{center}
\end{figure}

\begin{table}
\begin{center}
\caption{Comparison of \Fewbody\ with Table 5 of \citet{1983MNRAS.203.1107M}:  
      fraction of strong binary--binary 
      interactions that result in various outcomes.  In each binary--binary interaction the
      stars had equal masses and were assumed to be point particles, the binaries had equal 
      semimajor axes, and the eccentricities
      were drawn from a thermal distribution.  The data are normalized
      to 100 total scattering experiments.  The results are also shown graphically in Figure
      \ref{fig:mikkolacomp}.\label{tab:mikkolacomp}}
\begin{tabular}{rlrrrrrr}
\hline
$v_\infty/v_c$ &method  &undecided   &exchange     &triple       &single ionization &full ionization &total\\
\hline
$0.316$        &Mikkola &$4.7\pm1.2$ &$11.0\pm1.9$ &$24.3\pm2.8$ &$60.0\pm4.5$      &$0.0\pm0.0$     &300\\
               &\Fewbody &$1.6\pm0.2$ &$6.0\pm0.3$  &$22.8\pm0.7$ &$69.7\pm1.2$      &$0.0\pm0.0$     &5225\\
$0.500$        &Mikkola &$2.7\pm0.9$ &$7.3\pm1.6$  &$20.3\pm2.6$ &$69.7\pm4.8$      &$0.0\pm0.0$     &300\\
               &\Fewbody &$1.4\pm0.2$ &$6.7\pm0.4$  &$17.2\pm0.6$ &$74.7\pm1.3$      &$0.0\pm0.0$     &4366\\
$0.707$        &Mikkola &$0.7\pm0.5$ &$9.3\pm1.8$  &$11.7\pm2.0$ &$78.3\pm5.1$      &$0.0\pm0.0$     &300\\
               &\Fewbody &$0.5\pm0.1$ &$7.7\pm0.5$  &$6.8\pm0.5$  &$85.0\pm1.6$      &$0.0\pm0.0$     &3303\\
$0.866$        &Mikkola &$0.7\pm0.5$ &$16.7\pm2.4$ &$5.0\pm1.3$  &$77.7\pm5.1$      &$0.0\pm0.0$     &300\\
               &\Fewbody &$0.1\pm0.1$ &$8.2\pm0.5$  &$2.3\pm0.2$  &$89.4\pm1.5$      &$0.0\pm0.0$     &3827\\
$1.000$        &Mikkola &$0.0\pm0.0$ &$9.3\pm1.8$  &$1.7\pm0.7$  &$89.0\pm5.4$      &$0.0\pm0.0$     &300\\
               &\Fewbody &$0.1\pm0.0$ &$6.4\pm0.4$  &$0.6\pm0.1$  &$92.7\pm1.6$      &$0.2\pm0.1$     &3499\\
$1.225$        &Mikkola &$0.0\pm0.0$ &$8.3\pm1.7$  &$0.7\pm0.5$  &$73.7\pm5.0$      &$17.3\pm2.4$    &300\\
               &\Fewbody &$0.0\pm0.0$ &$4.6\pm0.3$  &$0.2\pm0.1$  &$92.1\pm1.5$      &$3.1\pm0.3$     &3969\\
\hline
\end{tabular}
\end{center}
\end{table}

\clearpage
\begin{figure}
  \begin{center}
    \includegraphics[width=0.48\textwidth]{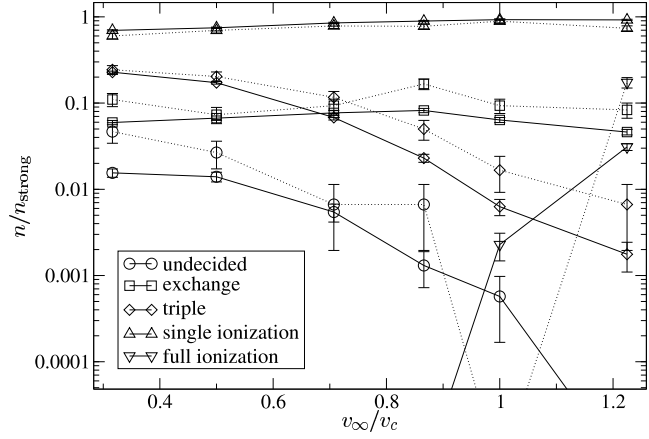}
    \caption{Comparison of \Fewbody\ (solid lines) with Table 5 of 
      \citet{1983MNRAS.203.1107M} (dotted lines): 
      fraction of strong binary--binary 
      interactions that result in various outcomes.  In each binary--binary interaction the
      stars had equal masses and were assumed to be point particles, the binaries had equal 
      semimajor axes, and the eccentricities
      were drawn from a thermal distribution.  Circles represent outcomes that were undecided
      after a preset maximum computation time, squares represent exchanges, diamonds
      represent stable hierarchical triples, upward-pointing triangles represent outcomes that resulted
      in one binary being disrupted, and downward-pointing triangles represent outcomes that resulted
      in both binaries being disrupted.  The solid lines represent \Fewbody\ data, while the dotted
      lines represent data from \citet{1983MNRAS.203.1107M}.  The results
      are also presented in Table \ref{tab:mikkolacomp}.\label{fig:mikkolacomp}}
  \end{center}
\end{figure}

\begin{figure}
  \begin{center}
    \includegraphics[width=0.48\textwidth]{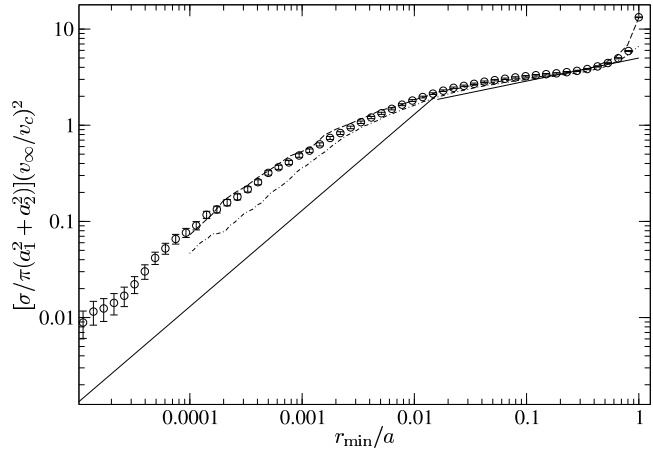}
    \caption{Comparison of \Fewbody\ with Figure 4 of \citet{1996MNRAS.281..830B}: cumulative
      cross section for the distance of closest approach in binary--binary scattering
      for the equal-mass, zero-eccentricity, equal-semimajor-axis case.  The stars are
      assumed to be point particles and $v_\infty/v_c=0.25$.  A total of 
      $1.5\times10^4$ scattering experiments were used to create this figure.  
      The broken power-law is the best fit given by \citet{1996MNRAS.281..830B}
      to their original results, while the dot-dash curve is Sigurdsson's recalculation.
      The dashed curve shows the results obtained using \Starlab.  There is a clear
      discrepancy between \Fewbody\ and \citet{1996MNRAS.281..830B}, and even the 
      recalculation.  However, \Fewbody\ and \Starlab\ agree quite well.\label{fig:baconcomp}}
  \end{center}
\end{figure}

\clearpage
\begin{figure}
  \begin{center}
    \includegraphics[width=0.48\textwidth]{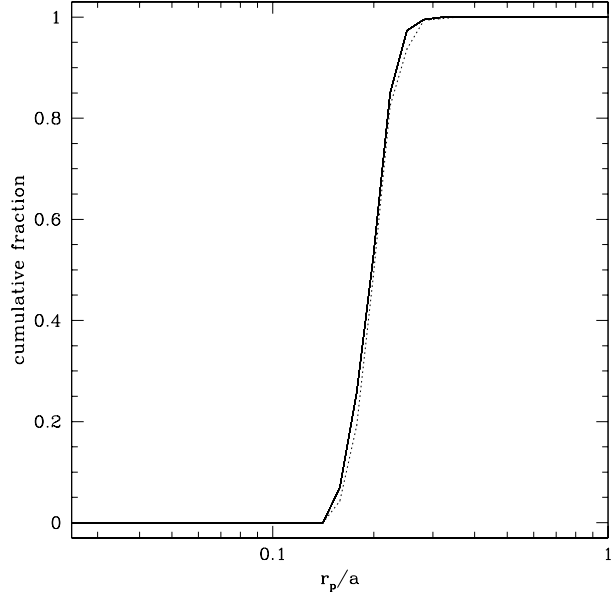}
    \caption{Comparison between {\em scatter3\/}, \Starlab's three-body scattering routine (solid line), 
      and {\em scatter\/}, its $N$-body scattering routine (dashed line).  Plotted is the cumulative
      fraction of binaries as a function of $r_p/a$, where
      $r_p$ is the pericentre distance of the merger binary, and $a$ is the initial
      binary semimajor axis.  For the binary--single runs,
      each star had mass $M_{\sun}$, radius $R_{\sun}$, the binary had
      semimajor axis $1\,\mbox{AU}$ and $e=0$, and $v_\infty=10\,\mbox{km/s}$.  In the 
      binary--binary runs, the binary representing the single star had a secondary 
      of mass $10^{-5}\,M_{\sun}$, semimajor axis of $20\,\mbox{AU}$, and $e=0$.
      The agreement between the two methods is excellent, and in either case,
      $r_p$ is strongly concentrated between $0.15\,\mbox{AU}$ and $0.3\,\mbox{AU}$.\label{fig:cfperi}}
  \end{center}
\end{figure}

\begin{figure}
  \begin{center}
    \includegraphics[width=0.45\textwidth]{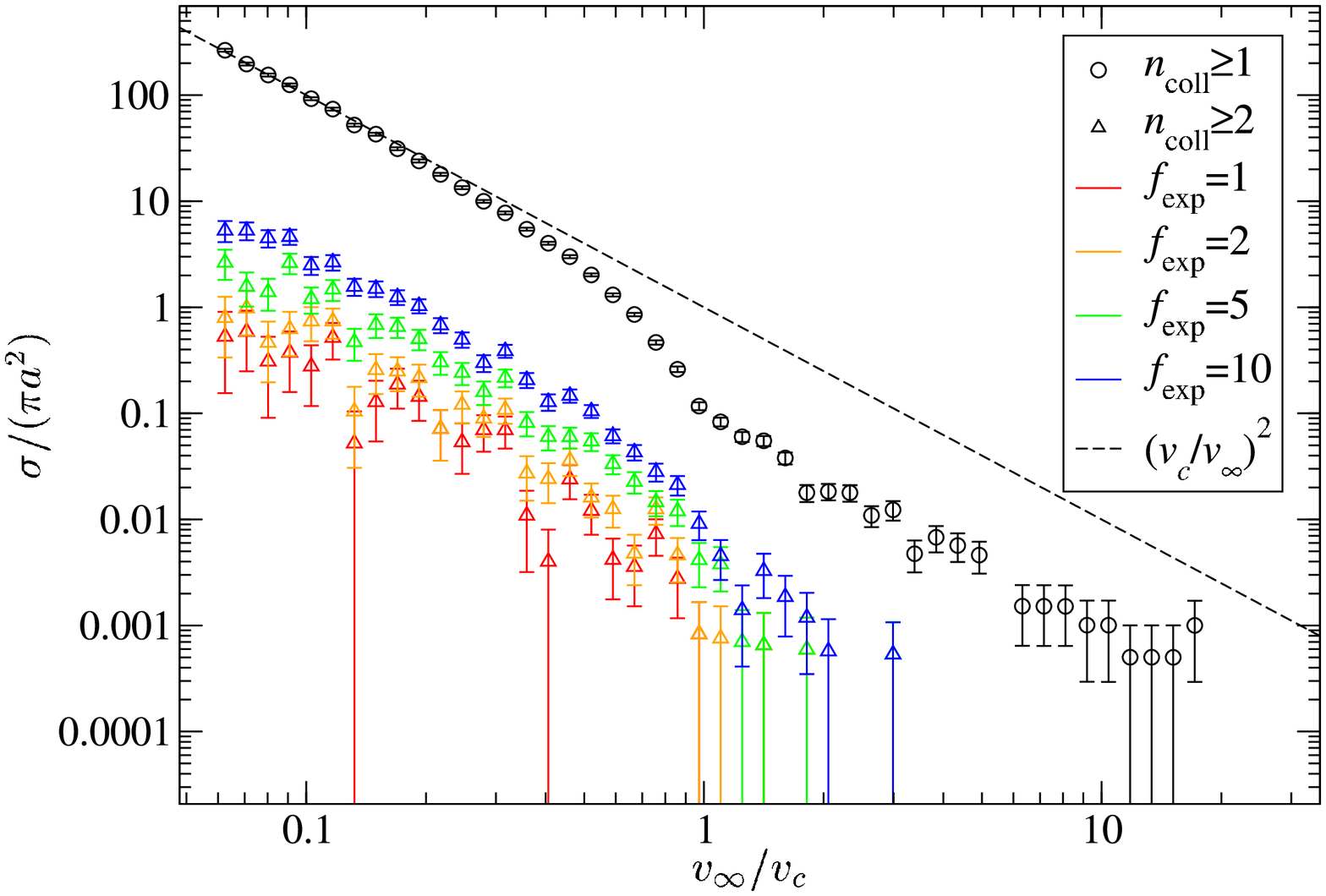}
    \hskip0.05\textwidth
    \includegraphics[width=0.45\textwidth]{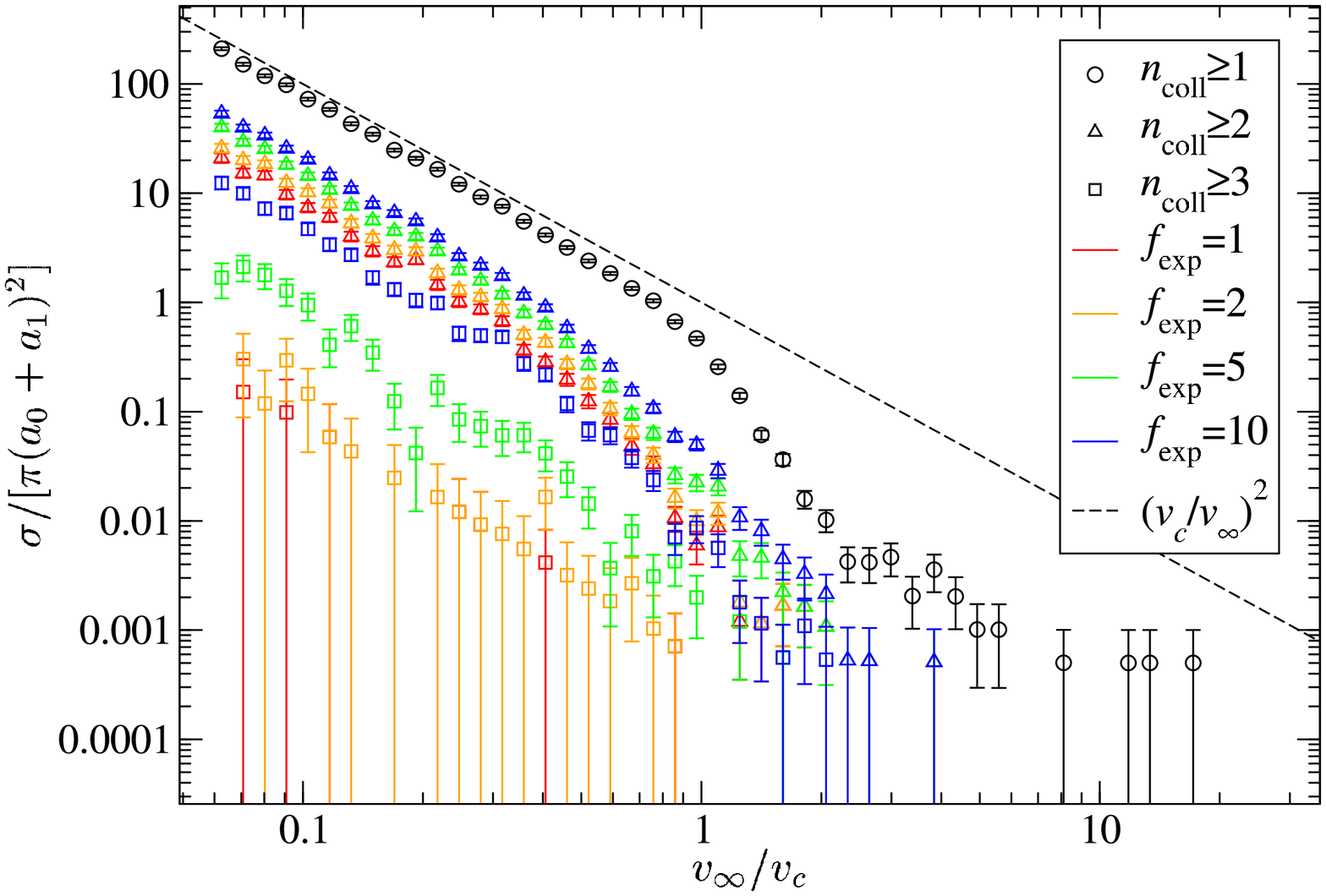}
    \caption{Cross section for physical collisions in binary--single (left)
      and binary--binary (right) scattering as a function of the relative velocity
      at infinity, for different values of the expansion parameter, $f_{\rm exp}$.
      Circles represent outcomes with one or more collisions; triangles, two or more; and
      squares, three or more.  Red represents runs with $f_{\rm exp}=1$; orange,
      $f_{\rm exp}=2$; green, $f_{\rm exp}=5$; and blue, $f_{\rm exp}=10$.
      In both experiments (binary--single and binary--binary), each star had mass
      $M_{\sun}$ and radius $R_{\sun}$, and each binary had semimajor axis 
      $a=1\,\mbox{AU}$ and eccentricity $e=0$.  The cross section decreases sharply at 
      the critical velocity, $v_c$,
      above which resonant scattering is forbidden. \label{fig:v_sigma}}
  \end{center}
\end{figure}

\clearpage
\begin{figure}
  \begin{center}
    \includegraphics[width=0.45\textwidth]{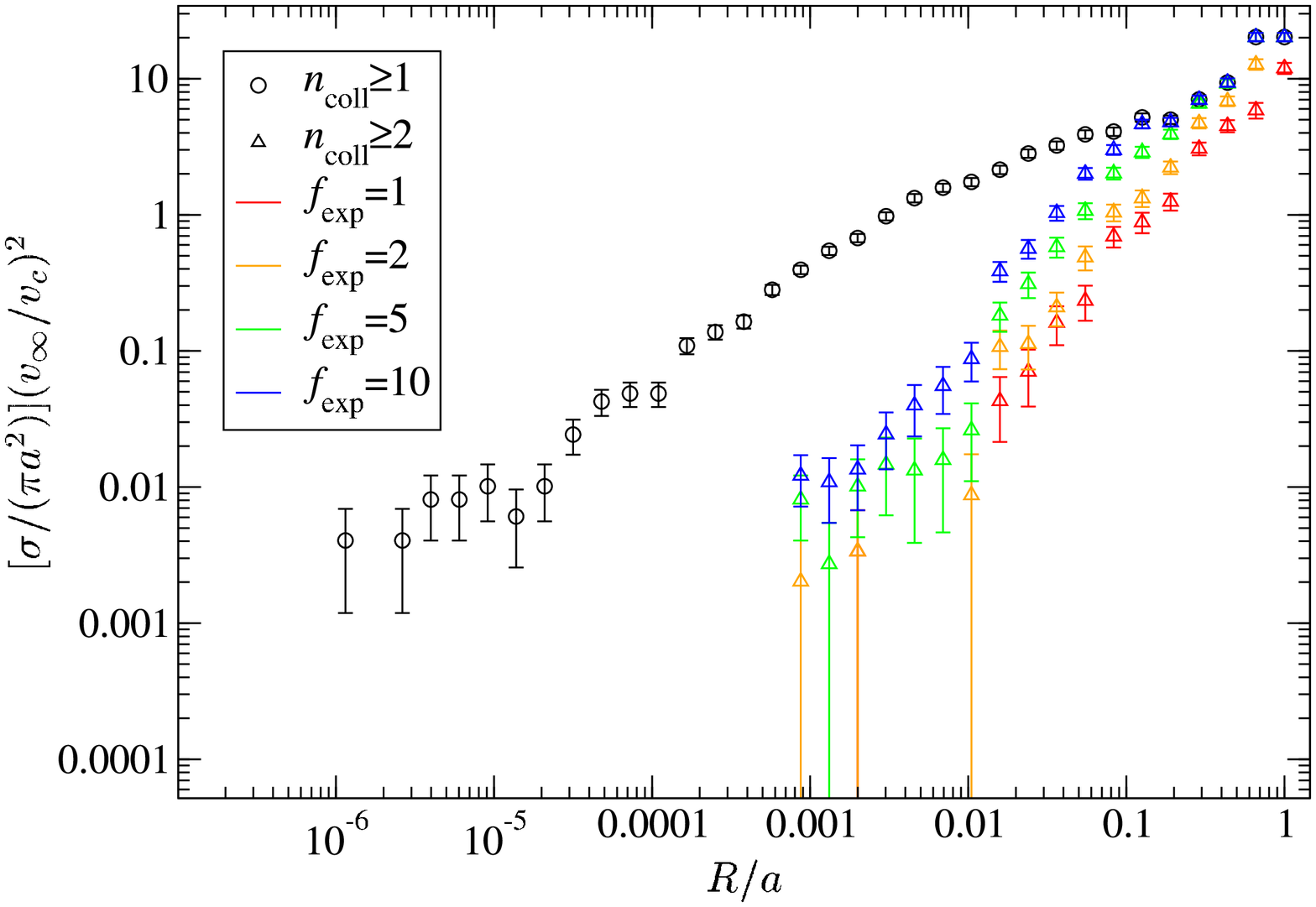}
    \hskip0.05\textwidth
    \includegraphics[width=0.45\textwidth]{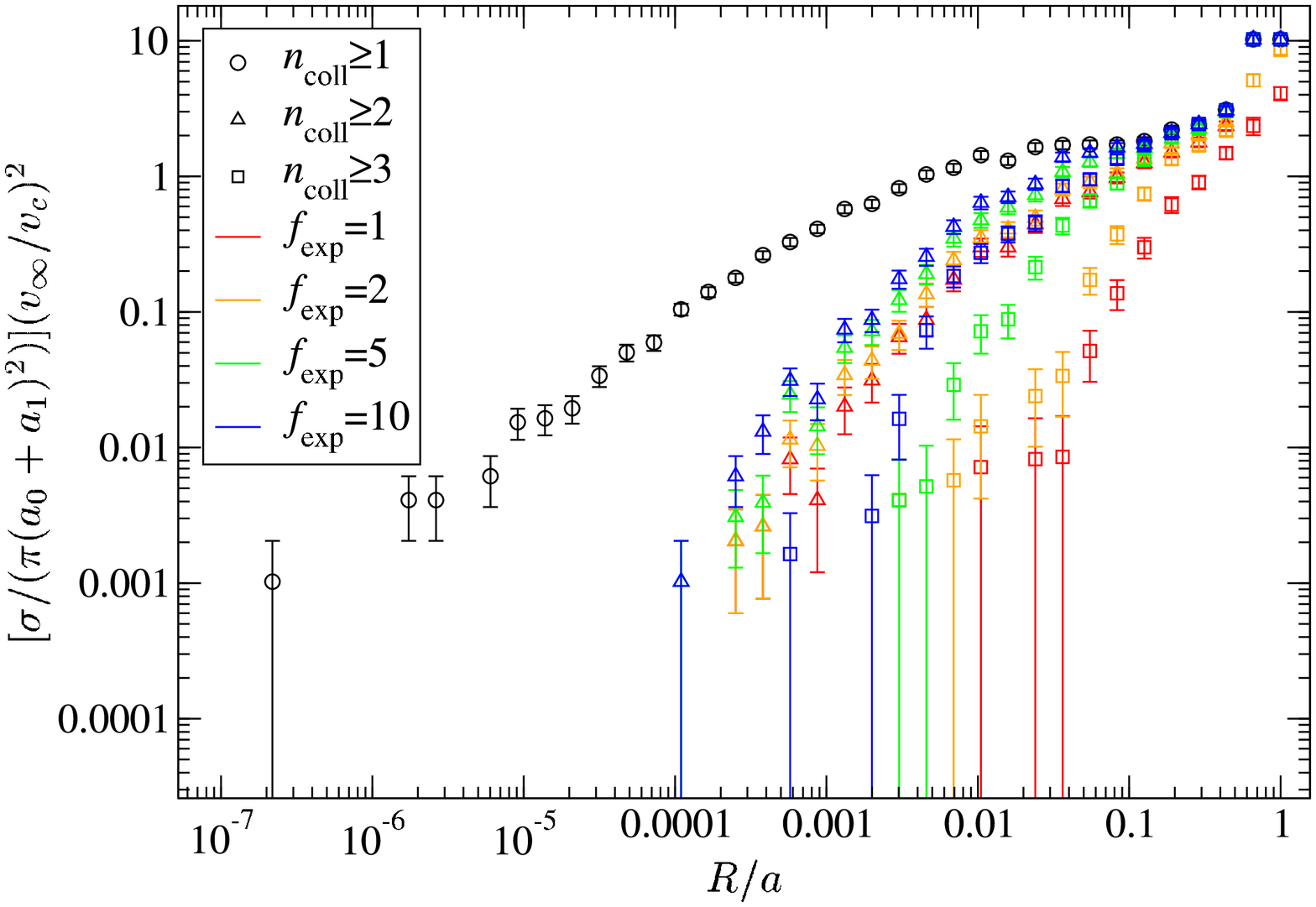}
    \caption{Normalized cross section for physical collisions in binary--single (left)
      and binary--binary (right) scattering as a function of the ratio of each star's
      radius to each binary's semimajor axis, $R/a$, for different values of 
      the expansion parameter, $f_{\rm exp}$.  Circles represent outcomes with one or more collisions; 
      triangles, two or more; and
      squares, three or more.  Red represents runs with $f_{\rm exp}=1$; orange,
      $f_{\rm exp}=2$; green, $f_{\rm exp}=5$; and blue, $f_{\rm exp}=10$.
      In both experiments 
      (binary--single and binary--binary), each star had mass $M_{\sun}$ and radius $R$,
      each binary had semimajor axis $a=1\,\mbox{AU}$ and eccentricity $e=0$, and the relative velocity
      at infinity was set to $v_\infty/v_c=0.1$.  Calculations were performed down
      to $R/a=10^{-9}$, but no collisions were found below $R/a\approx 10^{-6}$.
      \label{fig:r_sigma}}
  \end{center}
\end{figure}

\begin{figure}
  \begin{center}
    \includegraphics[width=0.45\textwidth]{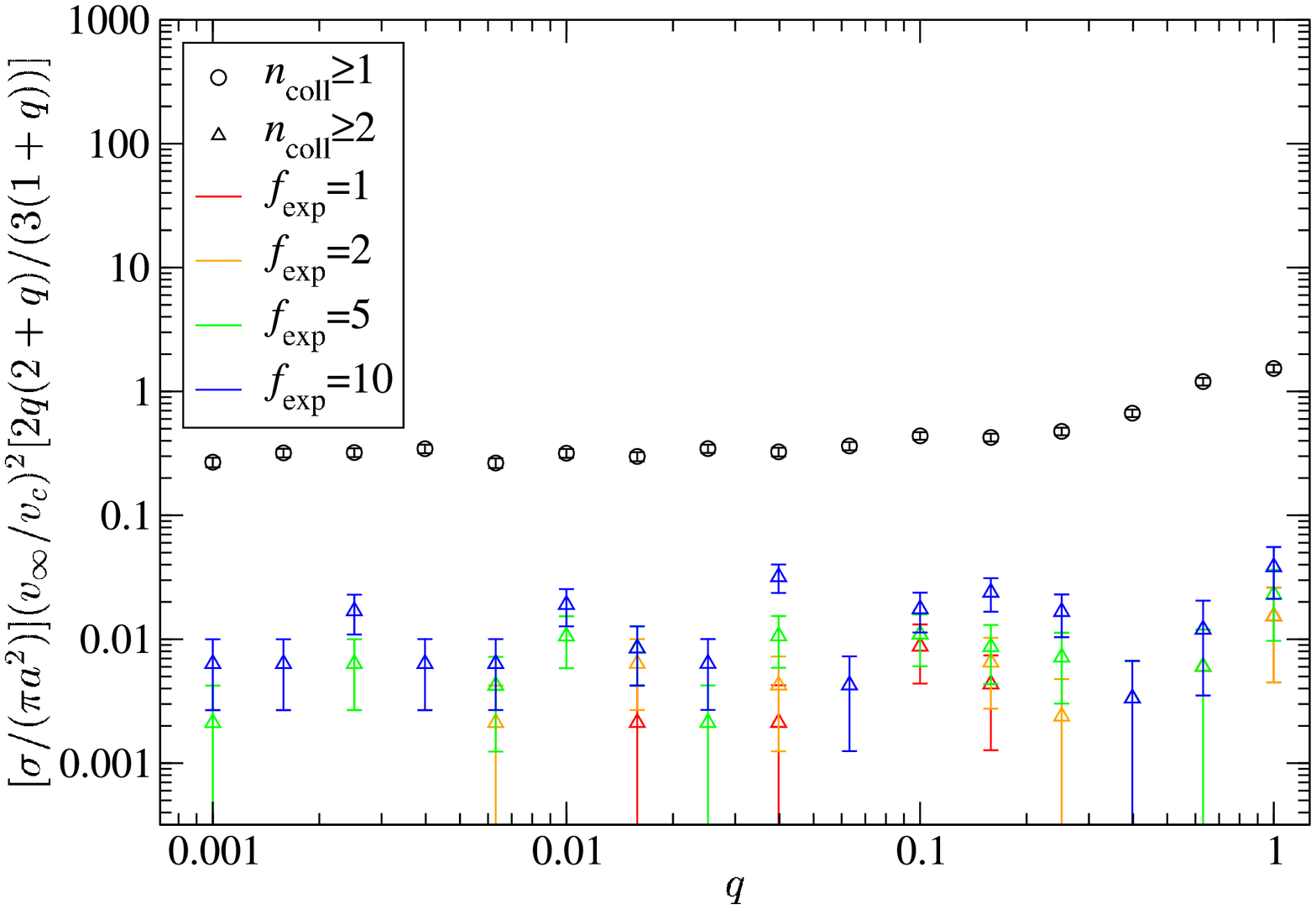}
    \hskip0.05\textwidth
    \includegraphics[width=0.45\textwidth]{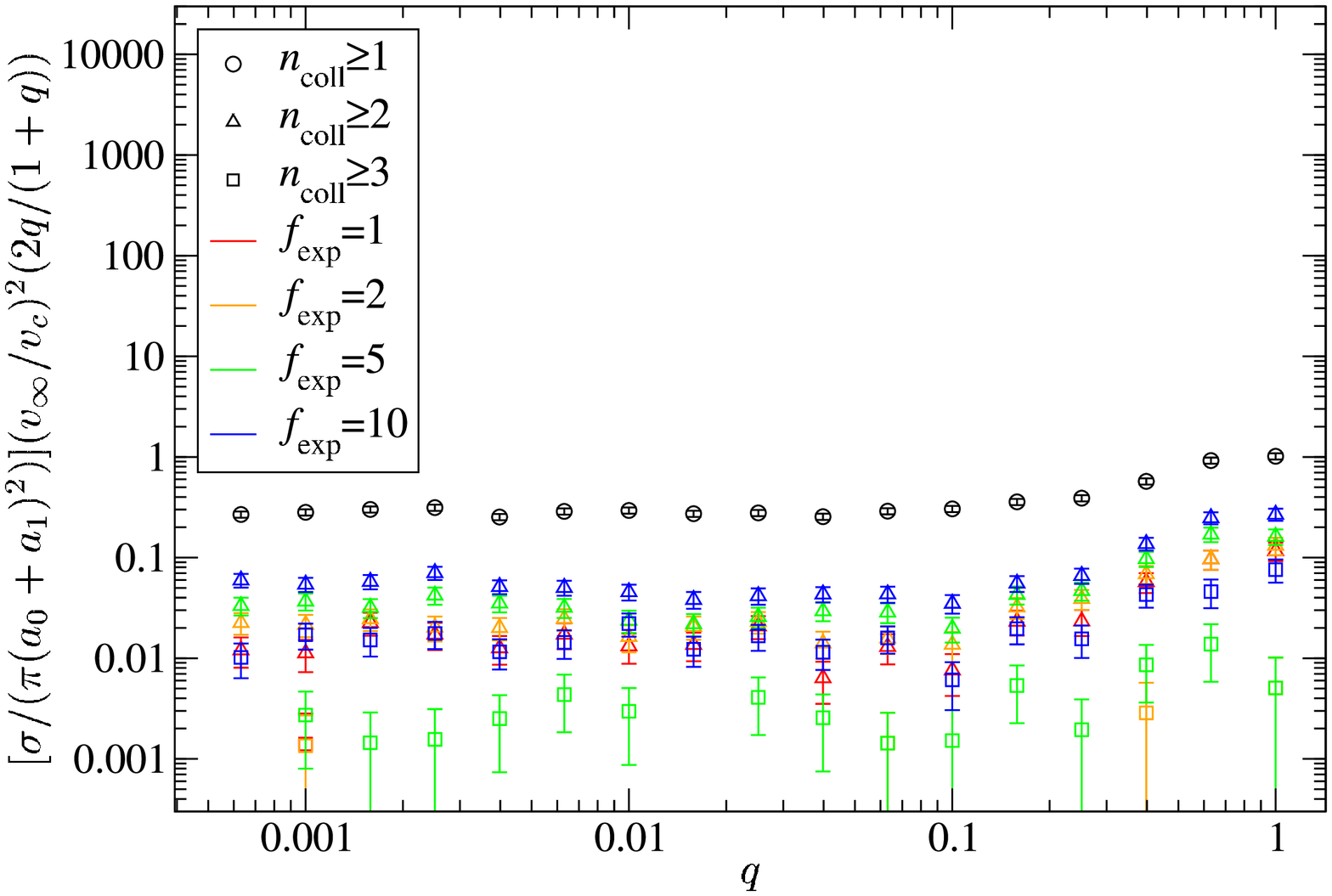}
    \caption{Normalized cross section for physical collisions in binary--single (left)
      and binary--binary (right) scattering as a function of mass ratio, $q$, 
      for different values of the expansion parameter, $f_{\rm exp}$.  
      Circles represent outcomes with one or more collisions; triangles, two or more; and
      squares, three or more.  Red represents runs with $f_{\rm exp}=1$; orange,
      $f_{\rm exp}=2$; green, $f_{\rm exp}=5$; and blue, $f_{\rm exp}=10$.
      In both experiments 
      (binary--single and binary--binary), each binary had one star with mass
      $M_{\sun}$ and the other with mass $q M_{\sun}$.  For the binary--single
      case, the incoming single star had mass $M_{\sun}$.  Each star had radius
      $R_{\sun}$, each binary had semimajor axis $a=1\,\mbox{AU}$ and
      eccentricity $e=0$, and the relative velocity at 
      infinity was set to $v_\infty/v_c=0.1$.  \label{fig:q_sigma}}
  \end{center}
\end{figure}

\clearpage
\begin{table}
\caption{Parameters of the binary--single runs, including the number of scattering 
interactions performed, $N$; the masses of the binary members, $m_{00}$ and $m_{01}$; 
the mass of the intruder, $m_1$; the binary semimajor axis, $a$; and the 
$n_{\rm coll} \geq 1$ cross section.\label{3body}}
\begin{center}
\vspace{0.2in}
\begin{tabular}{ccccccc}
\hline
run & $N$ & $m_{00}\,(M_{\sun})$ & $m_{01}\,(M_{\sun})$ & $m_1\,(M_{\sun})$ & $a\,(\mbox{AU})$ & $\displaystyle \frac{\sigma_{n_{\rm coll}\geq 1}}{\pi a^2}\left(\frac{v_\infty}{v_c}\right)^2$ \\
\hline
A005 & 15054 & 1.0 & 1.0 & 1.0 & 0.05 & $6.4\pm 0.1$ \\
A010 & 30228 & 1.0 & 1.0 & 1.0 & 0.1 & $5.72\pm 0.07$ \\
A020 & 15222 & 1.0 & 1.0 & 1.0 & 0.2 & $4.41\pm 0.08$ \\
A050 & 18158 & 1.0 & 1.0 & 1.0 & 0.5 & $2.87\pm 0.07$ \\
A100 & 37625 & 1.0 & 1.0 & 1.0 & 1.0 & $1.94\pm 0.04$ \\
A300 & 21427 & 1.0 & 1.0 & 1.0 & 3.0 & $0.68\pm 0.04$ \\
B005 & 17619 & 1.0 & 0.5 & 1.0 & 0.05 & $9.3\pm 0.2$ \\
B010 & 30408 & 1.0 & 0.5 & 1.0 & 0.1 & $7.3\pm 0.1$  \\
B020 & 17969 & 1.0 & 0.5 & 1.0 & 0.2 & $5.4\pm 0.1$ \\
B050 & 18676 & 1.0 & 0.5 & 1.0 & 0.5 & $3.0\pm 0.1$ \\
B100 & 39739 & 1.0 & 0.5 & 1.0 & 1.0 & $1.62\pm 0.05$ \\
B300 & 28544 & 1.0 & 0.5 & 1.0 & 3.0 & $0.54\pm 0.05$ \\
C005 & 17696 & 0.5 & 0.5 & 1.0 & 0.05 & $12\pm 2$ \\
C010 & 35791 & 0.5 & 0.5 & 1.0 & 0.1 & $9.3\pm 0.1$ \\
C020 & 18284 & 0.5 & 0.5 & 1.0 & 0.2 & $6.6\pm 0.2$ \\
C050 & 19467 & 0.5 & 0.5 & 1.0 & 0.5 & $3.3\pm 0.1$ \\
C100 & 49032 & 0.5 & 0.5 & 1.0 & 1.0 & $1.96\pm 0.07$ \\
C300 & 33464 & 0.5 & 0.5 & 1.0 & 3.0 & $0.58\pm 0.07$ \\
D005 & 12530 & 1.0 & 1.0 & 0.5 & 0.05 & $2.5\pm 0.1$ \\
D010 & 12555 & 1.0 & 1.0 & 0.5 & 0.1 & $2.2\pm 0.1$ \\
D020 & 12610 & 1.0 & 1.0 & 0.5 & 0.2 & $1.75\pm 0.1$ \\
D050 & 12780 & 1.0 & 1.0 & 0.5 & 0.5 & $1.16\pm 0.08$ \\
D100 & 15672 & 1.0 & 1.0 & 0.5 & 1.0 & $0.73\pm 0.07$ \\
D300 & 14185 & 1.0 & 1.0 & 0.5 & 3.0 & $0.26\pm 0.04$ \\
E005 & 60252 & 1.0 & 1.0 & 1.2 & 0.05 & $8.0\pm 0.2$ \\
E010 & 60504 & 1.0 & 1.0 & 1.2 & 0.1 & $6.9\pm 0.2$ \\
E020 & 61008 & 1.0 & 1.0 & 1.2 & 0.2 & $5.4\pm 0.2$ \\
E050 & 72947 & 1.0 & 1.0 & 1.2 & 0.5 & $3.5\pm 0.1$ \\
E100 & 75894 & 1.0 & 1.0 & 1.2 & 1.0 & $2.2\pm 0.1$ \\
E300 & 100200 & 1.0 & 1.0 & 1.2 & 3.0 & $0.80 \pm 0.05$ \\
\hline
\end{tabular}
\end{center}
\end{table}

\begin{figure}
\begin{center}
\includegraphics[width=0.48\textwidth]{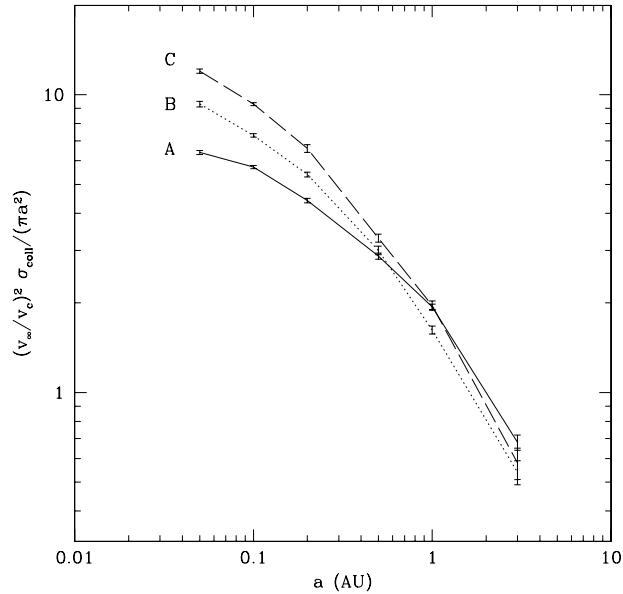}
\caption{\label{fig:abc}The normalized $n_{\rm coll}\geq 1$ cross section as a 
  function of initial semimajor axis for runs A, B and C.}
\end{center}
\end{figure}

\clearpage
\begin{figure}
\begin{center}
\includegraphics[width=0.48\textwidth]{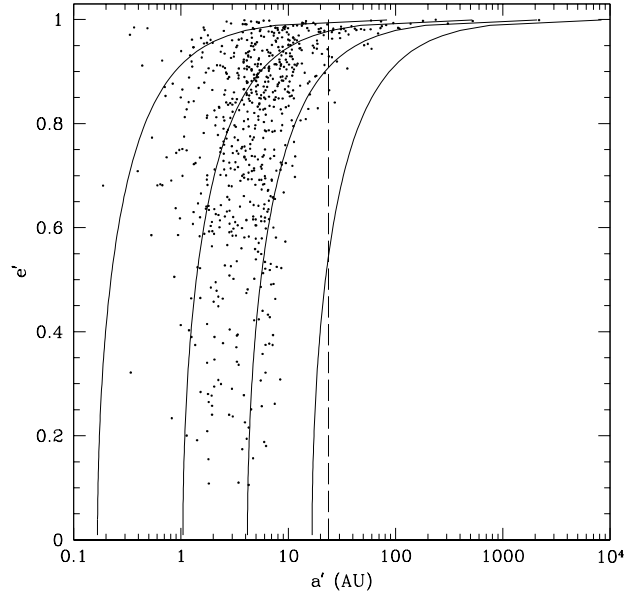}
\caption{Distribution of the semimajor axes and eccentricities of the $\sim 700$ merger binaries formed 
  in run A300. The vertical dashed line is the hard-soft boundary for field stars of 
  mass $1.0\,M_{\sun}$ with one-dimensional velocity dispersion $10\,\mbox{km/s}$. 
  The solid curves represent constant angular momenta $J/J_0=0.2$, 0.5, 1.0, and 2.0, where 
  $J_0$ is the total angular momentum of the system such that the pericentre of the initial 
  hyperbolic orbit is $1.0\,\mbox{AU}$. \label{mb1}}
\end{center}
\end{figure}

\begin{figure}
\begin{center}
\includegraphics[width=0.48\textwidth]{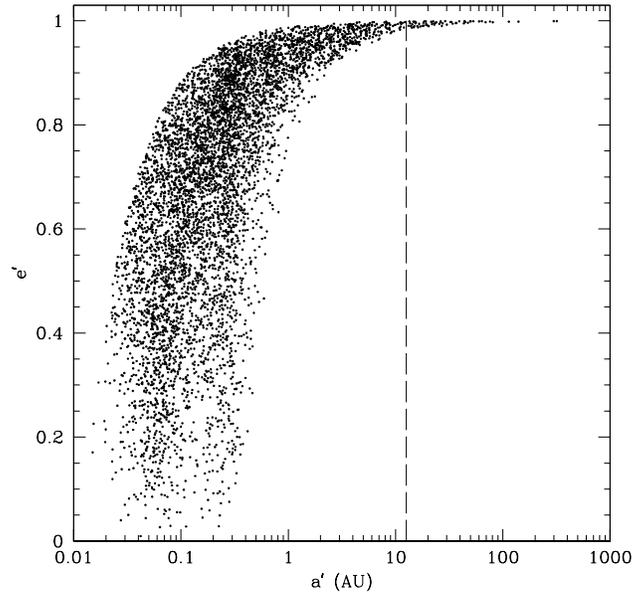}
\caption{Distribution of the semimajor axes and eccentricities of the $\sim 6000$ merger binaries 
  formed in run B005.  The vertical dashed line is the hard-soft boundary for field stars of 
  mass $1.0M_{\sun}$ with one-dimensional velocity dispersion $10\,\mbox{km/s}$.\label{mb2}}
\end{center}
\end{figure}

\clearpage
\begin{figure}
\begin{center}
\includegraphics[width=0.48\textwidth]{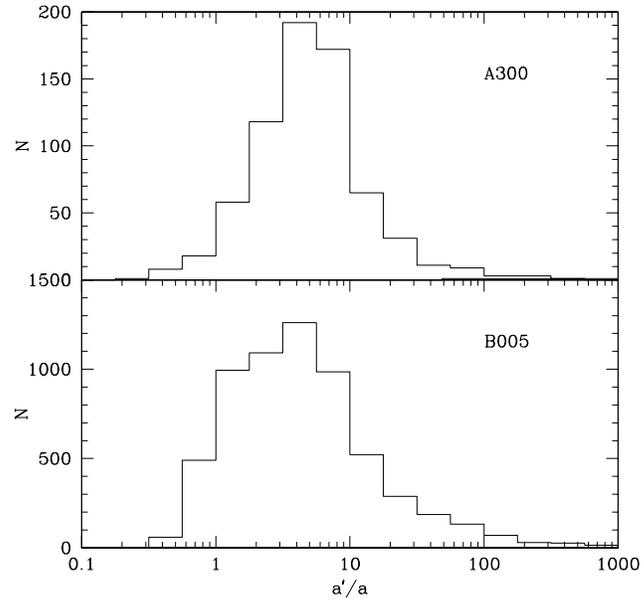}
\caption{Histograms of semimajor axes of the merger binaries formed in runs A300 and B005, relative
  to the initial binary semimajor axis. \label{hista} }
\end{center}
\end{figure}

\begin{figure}
\begin{center}
\includegraphics[width=0.48\textwidth]{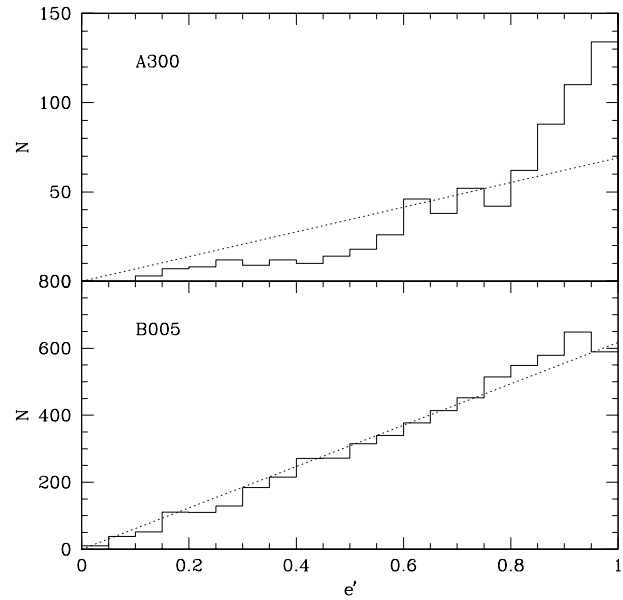}
\caption{Histograms of eccentricities of the merger binaries formed in runs A300 and B005. The 
dotted lines represent properly normalized thermal distributions. 
\label{histe}}
\end{center}
\end{figure}

\clearpage
\begin{figure}
\begin{center}
\includegraphics[width=0.48\textwidth]{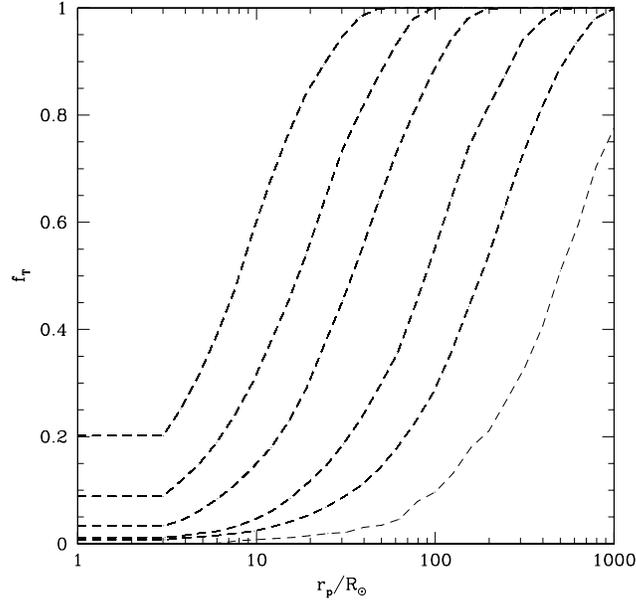}
\caption{Cumulative distribution of pericentre distance for the merger binaries formed
  in case A. The dashed lines, from left to right, correspond
  to $a=0.05$, 0.1, 0.2, 0.5, 1.0 and $3.0\,AU$ respectively.  Each curve is equivalent to 
  $f_T(R_{\rm cp})$, the fraction of triple mergers as a function 
  of the effective expanded radius of the first collision product, for a given $a$. \label{cumfrac}}
\end{center}
\end{figure}

\begin{figure}
\begin{center}
\includegraphics[width=0.48\textwidth]{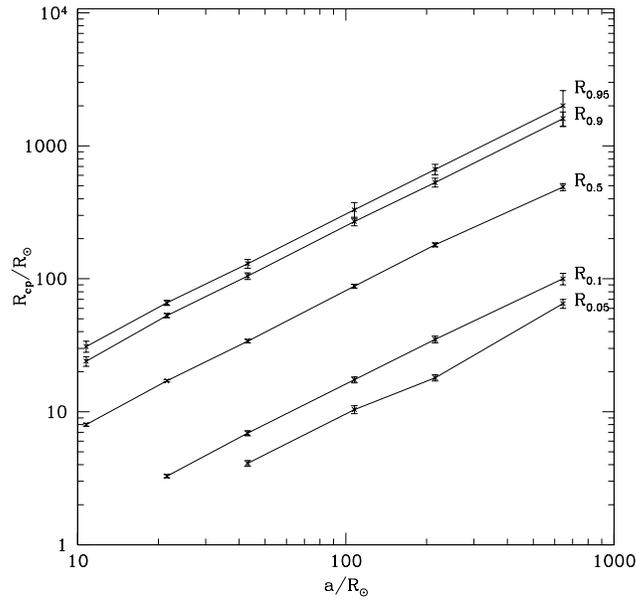}
\caption{$R_{0.95},R_{0.9},R_{0.5},R_{0.1}$ and $R_{0.05}$ as a function of $a$ for case A, 
where $R_f$ is the value of $R_{\rm cp}$ at which $f_T=f$. \label{rf}}
\end{center}
\end{figure}

\clearpage
\begin{figure}
\begin{center}
\includegraphics[width=0.48\textwidth]{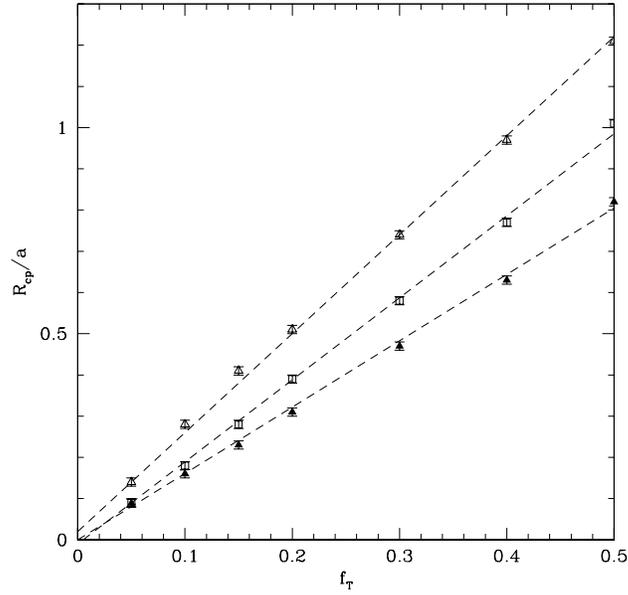}
\caption{Dependence of $R_f/a$ on $f=f_T$, with $f\leq 0.5$, for three different 
mass combinations.  Solid triangles, open squares, and open triangles correspond 
to cases A, B, and C, respectively. \label{raf}}
\end{center}
\end{figure}

\begin{table}
\begin{center}
\caption{Fits for $C$ in eq.(\ref{eqn:raf}) for the different mass combinations considered. \label{coef} }
\vspace{0.2in}
\begin{tabular}{ccccc}
\hline
Case & $m_{00}\,(M_{\sun})$ & $m_{01}\,(M_{\sun})$ & $m_1\,(M_{\sun})$ & $C$ \\
\hline
A &1.0 & 1.0 & 1.0 & 1.61$\pm$0.01 \\
B & 1.0 & 0.5 & 1.0 & 1.99$\pm$0.01 \\
C & 0.5 & 0.5 & 1.0 & 2.40$\pm$0.02 \\
D & 1.0 & 1.0 & 0.5 & 1.64$\pm$0.02 \\
E & 1.0 & 1.0 & 1.2 & 1.78$\pm$0.02 \\
\hline
\end{tabular}
\end{center}
\end{table}

\begin{table}
\begin{center}
\caption{Parameters of the binary--binary scattering experiments, 
  including the mass of each star, $m_{ij}$, the
  semimajor axis of each binary, $a_i$, and the normalized cross sections
  for strong interactions and at least one collision to occur.\label{4body}}
\vspace{0.2in}
\begin{tabular}{ccccccccc}
\hline
run & $m_{00}\,(M_{\sun})$ & $m_{01}\,(M_{\sun})$ & $m_{10}\,(M_{\sun})$ & $m_{11}\,(M_{\sun})$ & $a_0\,(\mbox{AU})$ & $a_1\,(\mbox{AU})$  & $\displaystyle \frac{\sigma_{\rm strong}}{\pi (a_0+a_1)^2}\left(\frac{v_\infty}{v_c}\right)^2$ & $\displaystyle \frac{\sigma_{n_{\rm coll}\geq 1}}{\pi (a_0+a_1)^2}\left(\frac{v_\infty}{v_c}\right)^2$ \\
\hline
I & 1.0 & 1.0 & 1.0 & 1.0 & 1.0 & 1.0 & $0.62\pm 0.13$ & $0.12\pm 0.07$ \\
II & 1.0 & 1.0 & 1.0 & 1.0 & 0.1 & 0.1 & $0.09\pm 0.03$ & $0.04\pm 0.02$ \\
III & 1.0 & 1.0 & 1.0 & 1.0 & 1.0 & 0.1 & $0.090\pm 0.003$ & $0.0020\pm 0.0005$ \\
IV & 1.0 & 0.5 & 1.0 & 0.5 & 0.1 & 0.1 & $0.12\pm 0.03$ & $0.0099\pm 0.0004$ \\
\hline
\end{tabular}
\end{center}
\end{table}

\clearpage
\begin{figure}
\begin{center}
\includegraphics[width=0.48\textwidth]{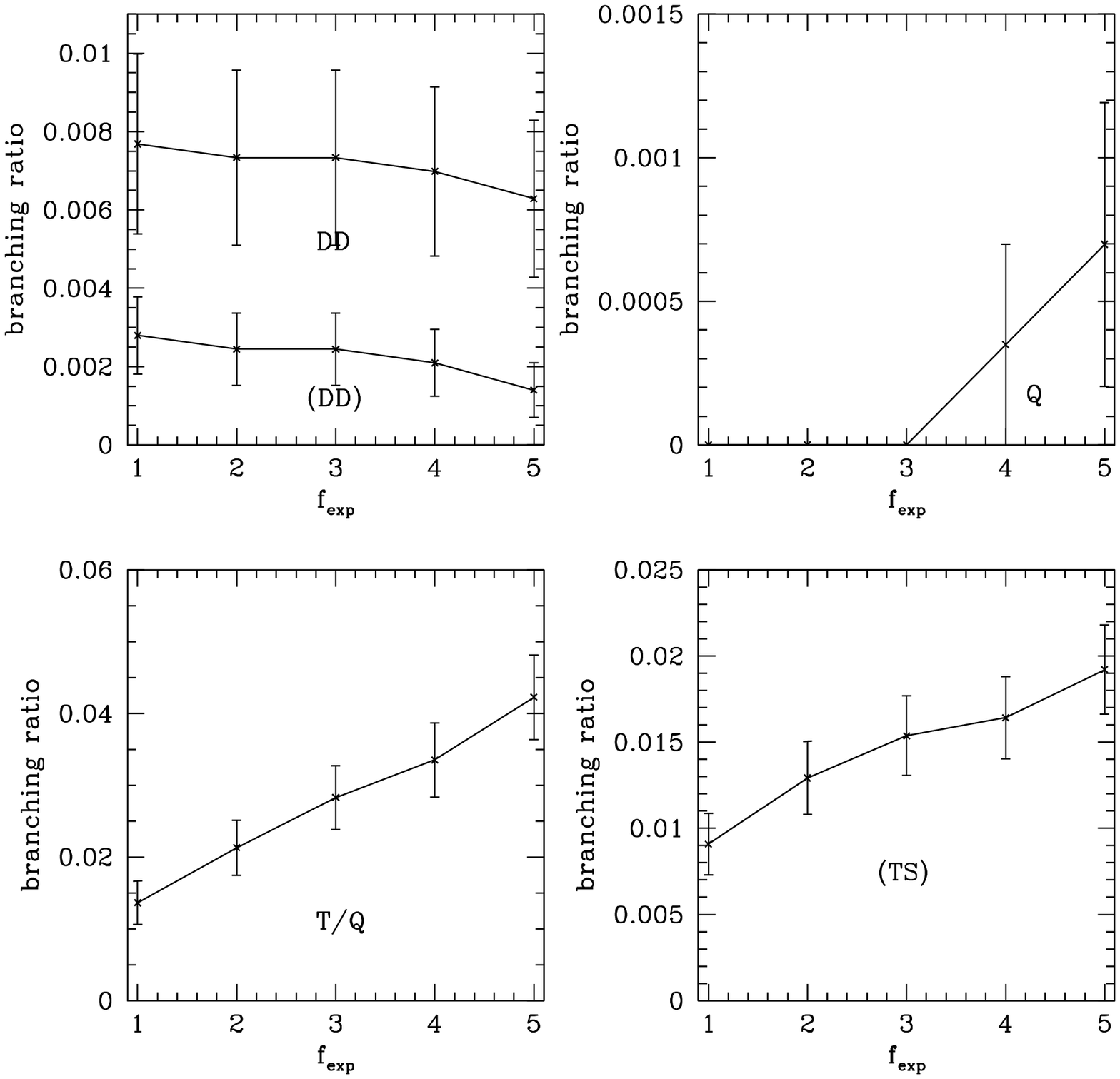}
\caption{Branching ratios for various outcomes involving collisions in run I, as functions
  of the expansion factor.\label{br1}}
\end{center}
\end{figure}

\begin{figure}
\begin{center}
\includegraphics[width=0.48\textwidth]{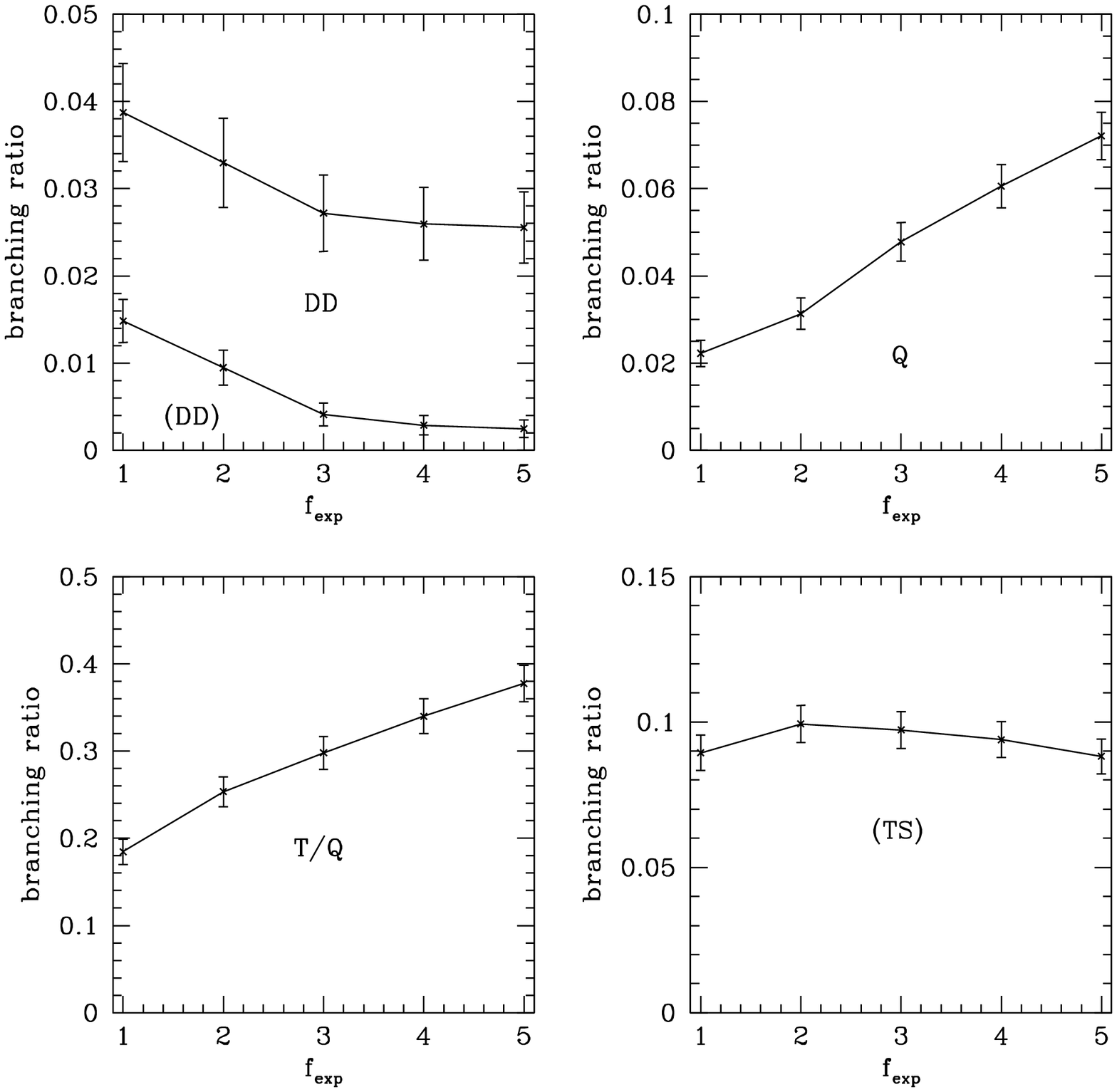}
\caption{Branching ratios for various outcomes involving collisions in run II, as functions
  of the expansion factor.\label{br2}}
\end{center}
\end{figure}

\clearpage
\begin{figure}
\begin{center}
\includegraphics[width=0.48\textwidth]{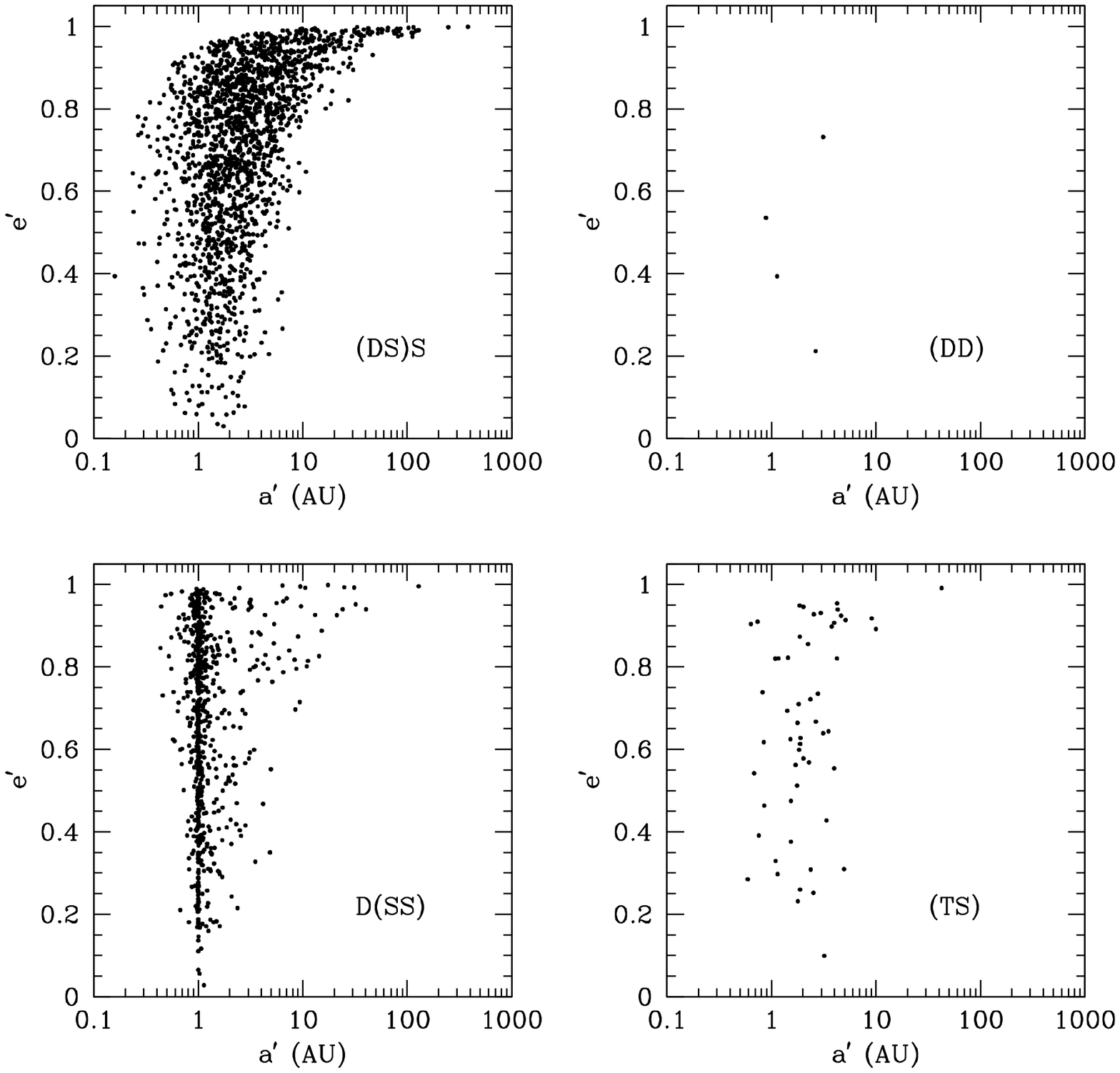}
\caption{Orbital parameters of four kinds of binaries formed in run I,
  with $f_{\rm exp}=5$.\label{ae1_5}}
\end{center}
\end{figure}

\begin{figure}
\begin{center}
\includegraphics[width=0.48\textwidth]{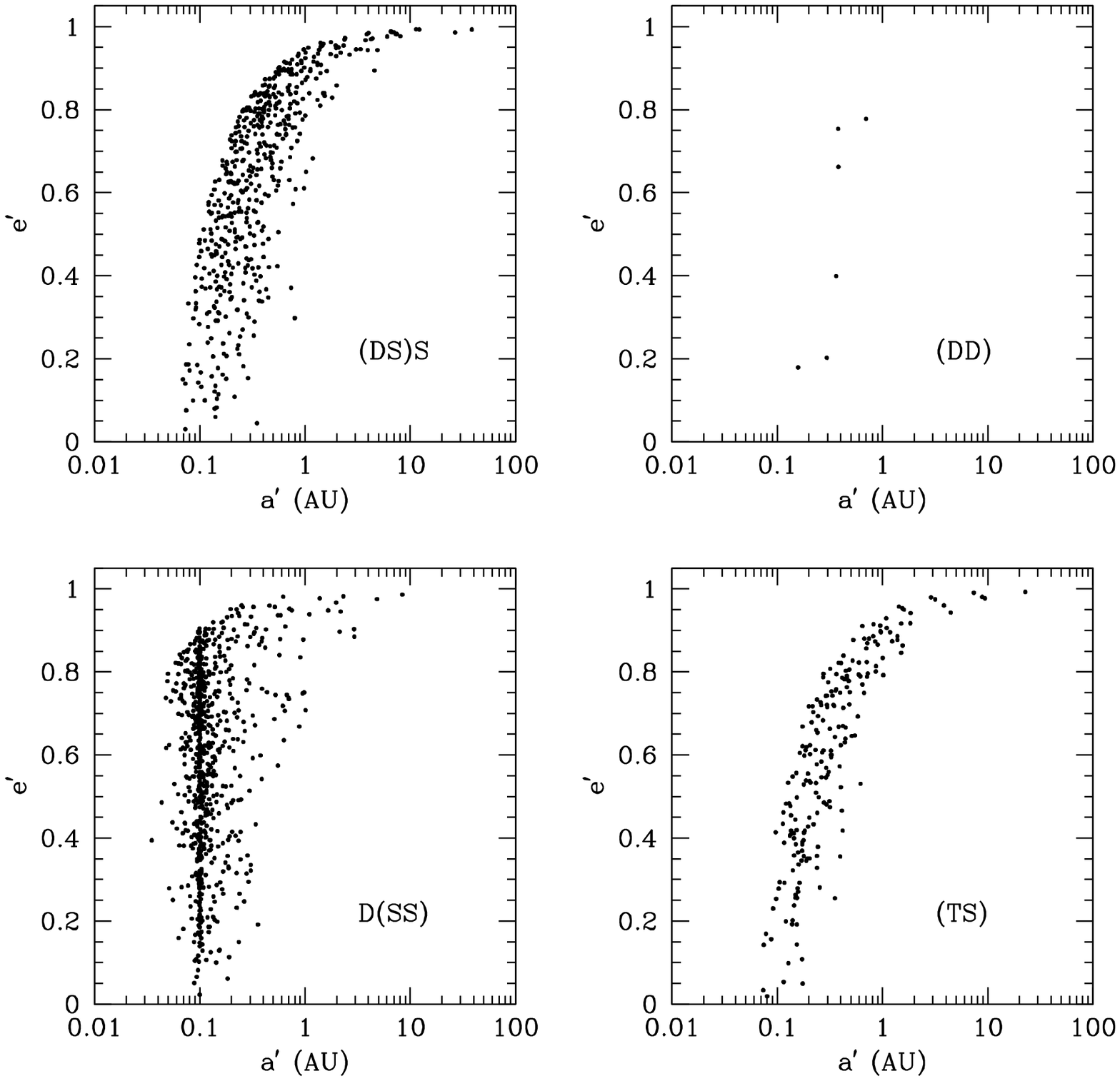}
\caption{Orbital parameters of four kinds of binaries formed in run II,
  with $f_{\rm exp}=5$.\label{ae2_5}}
\end{center}
\end{figure}

\begin{table}
\begin{center}
\caption{\label{fit}Linear fits for the branching ratio of T/Q($>2M_{\sun}$) as a function
  of $f_{\rm exp}$, where $f_{{\rm T/Q}(>2M_\odot)} = A f_{\rm exp}+B$.  Also shown in the last column
  is the normalized cross section for the formation of triple-star/quadruple-star mergers with masses
  $>2M_{\sun}$ for $f_{\rm exp}=5$.}
\vspace{0.2in}
\begin{tabular}{cccc}
\hline
run & $A$ & $B$ & $\displaystyle \frac{\sigma_{{\rm T/Q}(>2M_\odot)}(f_{\rm exp}=5)}{\pi (a_0+a_1)^2} \left(\frac{v_\infty}{v_c}\right)^2$ \\
\hline
I & 0.007 & 0.007 & $0.011\pm 0.002$ \\
II & 0.0485 & 0.14 & $0.035\pm 0.003$ \\
III & 0.0182 & 0.024 & $0.00041\pm 0.00005$ \\
IV & 0.0249 & 0.0616 & $0.034\pm 0.006$ \\
\hline
\end{tabular}
\end{center}
\end{table}

\end{document}